\documentclass[reqno]{amsart}
\usepackage{comment}
\usepackage{amssymb,latexsym,upref}

\usepackage{finalT}

\numberwithin{equation}{section}

\input free.def

\begin{document}

\title[Solutions to the relativistic Euler equations with a vacuum boundary]{On the existence of solutions to the relativistic Euler equations in $2$
spacetime dimensions with a vacuum boundary}
\author[T.A. Oliynyk]{Todd A. Oliynyk}
\address{School of Mathematical Sciences\\
Monash University, VIC 3800\\
Australia}
\email{todd.oliynyk@sci.monash.edu.au}

\begin{abstract}
We prove the existence of a wide class of solutions to the isentropic relativistic Euler equations in $2$ spacetime dimensions
with an equation
of state of the form $p=K\rho^2$ that have a fluid vacuum boundary. Near the fluid vacuum boundary, the sound speeds
for these solutions are monotonically decreasing, approaching zero where the density vanishes. Moreover, the
fluid acceleration is finite and bounded away from zero as the
fluid vacuum boundary is approached. The existence results of this article also generalize in a straightforward manner to equations of state
of the form $p=K\rho^\frac{\gamma+1}{\gamma}$ with
$\gamma > 0$.
\end{abstract}

\maketitle

\sect{intro}{Introduction}

Letting
\eqn{metric}{
g = g_{\mu \nu} dx^\mu dx^\nu \qquad \mu,\nu = 0,1
}
denote a flat Lorentz metric of signature $(+,-)$ on a 2 dimensional manifold $M \subset \Rbb^{2}$,
the relativistic Euler equations are
\leqn{eul11}{
\nabla_\mu T^{\mu\nu} = 0
}
where
\eqn{eul12}{
T^{\mu \nu} = (\rho + p)v^\mu v^\nu - p g^{\mu \nu} \AND g_{\mu\nu}v^\mu v^\nu = 1.
}
Here, $v^\mu$ is the fluid two-velocity, $\rho$ the proper energy density of the fluid, and $p$ the pressure.
Projecting \eqref{eul11} into the subspaces parallel and orthogonal to $v^\mu$ yields
the following well known form of the Euler equations
\lalign{eul13}{
v^\mu \nabla_\mu \rho + (\rho+p)\nabla_\mu v^\mu & =  0, \label{eul13.1}\\
(\rho + p)v^\mu \nabla_\mu v^\nu - h^{\mu\nu}\nabla_\mu p & = 0 \label{eul13.2}
}
where
\leqn{eul14}{
h_{\mu\nu} = g_{\mu\nu} - v_\mu v_\nu \qquad (v_\mu = g_{\mu\nu} v^\nu)
}
is the induced negative definite metric on the subspace orthogonal to $v^\mu$.
In this article, we will be primarily concerned
with fluids with an insentropic equation of state of the form
\leqn{fwexA1}{
p = K \rho^{\frac{\gamma+1}{\gamma}} \qquad K,\gamma >0,
}
and in particular $\gamma =1$.

The main aim of this article is to construct solutions to the Euler equations that contain a vacuum region; that is
solutions for which $\rho$ vanishes, and the fluid vacuum boundary is located where $\rho=0$.
The problem of existence of solutions to the Euler equations with a vacuum region for equations of state for
which the pressure and proper energy density simultaneously vanish has been studied by a number of authors. The main difficulty in establishing
existence is due to the fact that the Euler equations  \eqref{eul13.1}-\eqref{eul13.2} become degenerate as the fluid
vacuum boundary is approached. This is due to the vanishing of the the quantity $\rho + p$ at the fluid vacuum boundary.
Because of this, standard techniques from
hyperbolic PDE theory do not apply. The first general existence result where this problem was overcome is in
\cite{Mak}. In this work, the existence of gravitating, non-relativistic fluid bodies with compactly supported densities in 4 spacetime dimensions was established (see \cite{Ren} and \cite{LU} for the relativistic case);
the technique employed involved a special choice of variables to regularize the Euler equations, and works in all dimensions
with or without coupling to gravity.
However, the type of fluid solutions obtained by this method have, when coupled to gravity, freely falling boundaries,
and hence, do not include static or expanding fluid bodies. This is because the fluid acceleration
\leqn{accel0}{
a^\nu = v^\mu \nabla_\mu v^\nu
}
of the solutions from \cite{Mak} vanish at the fluid vacuum boundary, and so there is no outward force to counteract
the gravitation force which leads to collapse of the body. Consequently, to have a class of solutions
which include the full physical range including static and expanding fluid bodies, it is necessary to
establish the existence of
solutions to the Euler equations for which the acceleration \eqref{accel0} is non-zero at the fluid vacuum boundary.
Due to the finite propagation speed of the Euler equations, it is enough to consider the existence problem in a neighborhood
of the fluid vacuum boundary. Away from the boundary where $\rho+p>0$, standard symmetric hyperbolic techniques can
be used.

In this article, we establish the existence of a wide class of solutions in 2 spacetime dimensions for which
the acceleration is non-zero at the boundary. The approach we take to establish existence relies on geometric
arguments and is inspired by the Lagrangian formulation of the relativistic Euler equations in \cite{Frie}, although
we use a different formulation of the relativistic Euler equations due to Frauendiener and Walton \cite{Frau,Walt}.
In the Frauendiener-Walton formulation, the normalized two-velocity $v^\mu$ and proper energy density $\rho$ are combined into a single vector $w^\mu$.
Due to the vector nature of this
formulation, it is possible to exploit the
full diffeormophism freedom available to fix the coordinates, and reduce the problem to that of two linear
wave equations one of which is regular while the other singular. Existence for
the regular wave equation is handled by standard PDE techniques while existence for the singular wave equation is obtained
by standard results on abstract wave equations.
Although the analysis in this article applies only to 2 spacetime dimensions, some of the arguments presented here do have
a counterpart in higher dimensions. However, we will not discuss these results here.

In 2 spacetime dimensions, the existence of solutions to the non-relativistic Euler equations with non-zero acceleration at the fluid vacuum boundary
has been previously established in the remarkable articles \cite{CK1D,JM}. In these articles, existence is proved using non-standard energy estimates
combined
with suitable approximation techniques. The arguments used are technical, involved, highly original, and quite different from
one another. We also note that the results in \cite{CK1D,JM} have been recently extended to 4 spacetime dimensions \cite{CK3D,JM3D} (see also \cite{CLS3D}).
At the moment,  the relationship between the solutions
of this article and those of \cite{CK1D,JM} is not clear because the solutions presented here are relativistic while those
from \cite{CK1D,JM} are non-relativistic. In order to understand the relationship, it would be necessary to
take the non-relativistic (i.e. $c\rightarrow \infty$) limit of our solutions. We will not do this here, but leave the
analysis of this limit to a separate article.

\bigskip

\noindent\textbf{Overview of the article:} In Section \ref{fw}, we review the Frauendiener-Walton formulation of
the isentropic Euler equations, and we use this formulation to show, in Section \ref{eulf}, that the Euler equations
are equivalent to the existence of a suitably normalized, commuting set of vector fields. This commuting set of
vector fields allows us to introduce adapted coordinates for the Lorentzian metric. As detailed in Section \ref{Ewave},
the flatness of the Lorentzian
metric then implies the equivalence of the Euler equations with a non-linear scalar wave equation. Conformal coordinates are introduced
in Section \ref{null} which reduce, as described in Section \ref{cwave}, the non-linear wave equation to a linear one.
In Section \ref{ccord}, it is shown that the existence of conformal coordinates is equivalent to the existence of suitable
solutions for a linear wave equation with singular coefficients. An existence and regularity theory for this type of wave
equation is developed in the Appendix \ref{wave}. The main result of this paper is contained in Section \ref{exist} where
the existence of solutions to the Euler equations with non-zero acceleration at the vacuum boundary is established, see
Theorem \ref{ethm} for a precise statement.   Finally, in Section \ref{exact}, a class of exact solutions with non-zero
acceleration at the vacuum boundary are described.

\sect{fw}{The Frauendiener-Walton formulation of the Euler equations}

In \cite{Frau,Walt}, Frauendiener and Walton independently showed that the isentropic Euler equations for a perfect fluid with an
equation of state of the form $p = p(\rho)$
can be written as\footnote{It is important to note that Frauendiener and Walton use opposite signature conventions for the metric $g$ and different notation
for the fluid variables. In this article, we use the notation and signature conventions of Frauendiener \cite{Frau}.}
\leqn{eul1}{
A_{\mu\nu}{}^\gamma \nabla_\gamma w^\nu = 0,
}
where $w^\nu$ is a timelike vector field with norm
\eqn{eul2}{
w^2 = w_\nu w^\nu > 0\qquad (w_\mu = g_{\mu\nu} w^\nu),
}
and
\leqn{eul3}{
A_{\mu\nu}{}^\gamma = \left(3 + \frac{1}{s^2}\right) \frac{w_\mu w_\nu}{w^2} w^\gamma - \delta^\gamma_\nu w_\mu
-\delta^\gamma_\mu w_\nu - w^\gamma g_{\mu\nu}.
}
We will refer to these equations as the \emph{Euler-Frauendiener-Walton (EFW)
equations}.

In the Frauendiener-Walton formulation, $s^2$ is a function of
\eqn{eul4}{
\zeta =\frac{1}{w},
}
where
\eqn{eul5}{
w= \sqrt{w^2}. }
An explicit formula for $s^2$ can be calculated in the following fashion (see \cite{Frau} for more details). First,
the pressure
$p=p(\zeta)$ is determined implicity by the equation
\leqn{eul6}{
\zeta = \zeta_0\Upsilon(p(\zeta)),
}
where
\leqn{eul7}{
\Upsilon(p) = \exp\left(\int_{p_0}^p \frac{d\tilde{p}}{\rho(\tilde{p})+\tilde{p}}\right)
}
is the Lichnerowicz index of the fluid. From this, $s^2$ can be calculated using
the formula
\leqn{eul8}{
\frac{1}{s^2} = \left(\frac{\zeta f'(\zeta)}{f(\zeta)}-3 \right)
}
where
\eqn{eul9}{
f(\zeta) = \zeta^3 p'(\zeta).
}
Additionally, the proper energy density $\rho$ and two-velocity
$v^\mu$ can be recovered from
\leqn{eul10}{
\rho = p(\zeta)-\frac{f(\zeta)}{\zeta^2} \AND v^\mu = \zeta w^\mu.
}
As shown in \cite{Frau}, and also \cite{Walt}, the triple $\{\rho,p,v^\mu\}$ determined from \eqref{eul1}, \eqref{eul6}, and \eqref{eul10} satisfy the relativistic
Euler equations \eqref{eul13.1}-\eqref{eul13.2}.

For the equations of states \eqref{fwexA1}, it is possible to explicitly determine the functional
form of $s^2=s^2(\zeta)$. To see this, we observe
that the Lichnerowicz index \eqref{eul7} is given by
\eqn{fwexA2}{
\Upsilon(p) = \exp\left(\int_{0}^p \frac{d\tilde{p}}{\rho(\tilde{p})+\tilde{p}}\right) = \bigl(1+K^{\frac{\gamma}{\gamma+1}}p^{\frac{1}{\gamma+1}}\bigr)^{\gamma+1}.
}
From this expression, it is clear
\eqn{fwexA3}{
p(\zeta) = \frac{1}{K^{\gamma}} \left(\left(\frac{\zeta}{\zeta_0}\right)^{\frac{1}{\gamma+1}}-1\right)^{\gamma+1}
}
solves \eqref{eul6}, and hence determines the pressure as a function of $\zeta$. Without loss of generality, we set $\zeta_0=1$ which gives
\eqn{fwexA4}{
p(\zeta) = \frac{1}{K^{\gamma}} \bigl(\zeta^{\frac{1}{\gamma+1}}-1\bigr)^{\gamma+1}.
}
Substituting this expression into \eqref{eul8} then yields
\leqn{fwexA5}{
s^2 = \frac{\gamma+1}{\gamma}\bigl(\zeta^{\frac{1}{\gamma+1}}-1\bigr).
}

For latter use, we note that by using \eqref{eul13.2} the fluid acceleration can be written as
\eqn{accel1}{
a^\nu  = \frac{h^{\mu\nu}}{\rho+p}\nabla_\mu p.
}
Since this vector is orthogonal to $v^\mu$, i.e. $v_\nu a^\nu = 0$, we can use the metric $h_{\mu\nu}$
to calculate the length of $a^\nu$ via the formula
\leqn{accel2}{
|a|_h := \sqrt{-h_{\mu\nu}a^\mu a^\nu} = \sqrt{-\frac{h^{\mu\nu}}{(\rho+p)^2}\nabla_\mu p \nabla_\nu p }\, .
}
Recalling that the square of the sound speed is given by
\leqn{sound}{
s^2 = \frac{dp}{d\rho} = \frac{K(\gamma+1)}{\gamma} \rho^{\frac{1}{\gamma}},
}
we can write \eqref{accel2} as
\leqn{accel3}{
|a|_h = \sqrt{-\frac{\gamma^2 h^{\mu\nu}}{(1+K\rho^{\frac{1}{\gamma}})^2}\nabla_\mu s^2 \nabla_\nu s^2 }
= \frac{\gamma}{(1+K\rho^{\frac{1}{\gamma}})}|\ed s^2|_h.
}
This shows that the acceleration will be non-zero at the fluid vacuum boundary if and only if
$|\ed s^2|_h > 0$ there, since $\rho$ vanishes at the boundary. It is also clear from this formula that
the fluid acceleration can only be non zero at the fluid vacuum boundary if the gradient of $s^2$ does
not vanish there. Since the proper energy density and hence the sound speed is decreasing to zero as the fluid
vacuum boundary is approached, it follows that for such solutions, $s^2$ must be monotonically decreasing
in the neighborhood of the boundary.

\sect{eulf}{A frame formulation for the EFW equations}

Since the metric $g=g_{\mu\nu} dx^\mu dx^\nu$ is flat,
we begin by introducing global Minkowskian coordinates\footnote{We note that this step of introducing
a global Minkowskian set of coordinates is not necessary.
We include it in order to make the following arguments more accessible to readers who may
not be familiar with a more abstract approach to fixing coordinates. A more abstract approach
becomes particularly important when dealing with non-flat metric that arise when coupling the fluid to
gravity. }
$(x^\mu)=(x^0,x^1)$ for which
\leqn{gcart}{
(g_{\mu\nu})=\begin{pmatrix} 1 & 0 \\ 0 & -1 \end{pmatrix}.
}
With this choice of coordinates, the Euler equations \eqref{eul13.1}-\eqref{eul13.2}
can be written explicitly
in terms of the proper energy density $\rho(x^0,x^1)$ and a function $v(x^0,x^1)$ that satisfy
\alin{eulcart}{
\frac{1}{\sqrt{1-v^2}}\del{0}\rho + \frac{v}{\sqrt{1-v^2}}\del{1}\rho
+ \bigl(\rho + K\rho^{\frac{\gamma+1}{\gamma}} \bigr)
\left( \del{0}\left(\frac{1}{\sqrt{1-v^2}}\right) +\del{1}
\left(\frac{v}{\sqrt{1-v^2}}\right) \right) & = 0 \\
\bigl(\rho + K\rho^{\frac{\gamma+1}{\gamma}} \bigr)\left(
\frac{1}{\sqrt{1-v^2}}\del{0}\left(\frac{v}{\sqrt{1-v^2}}\right) +
\frac{v}{\sqrt{1-v^2}}\del{1} \left(\frac{v}{\sqrt{1-v^2}}\right)
\right) \qquad \qquad & \\
+ K \frac{\gamma+1}{\gamma} \rho^{\frac{1}{\gamma}}\left(
\frac{v}{1-v^2}\del{0}\rho + \left(1+\frac{v^2}{1-v^2}\right)\del{1}\rho
\right) & = 0
}
with the two-velocity given by
\leqn{vcart}{
(v^\mu) = \frac{1}{\sqrt{1-v^2(x^0,x^1)}}\bigl( 1 , v(x^0,x^1)\bigr) .
}

Using \eqref{fwexA5} and \eqref{sound}, a short calculation shows that
\eqn{wcartA}{
w = \frac{1}{(1+K\rho^{\frac{1}{\gamma}})^{\gamma+1}}
}
from which we obtain
\leqn{wcartB}{
(w^\mu) = \frac{1}{\bigl(1+K\rho(x^0,x^1)^{\frac{1}{\gamma}}\bigr)^{\gamma+1}\sqrt{1-v(x^0,x^1)^2}}\bigl( 1 , v(x^0,x^1)\bigr)
}
by \eqref{vcart} and \eqref{eul10}. Letting
\eqn{A0def}{
A^0 = \begin{pmatrix}
\left[\begin{displaystyle}\left(3+\frac{1}{s^2}\right)\frac{(w^0)^2}{w^2}-3\end{displaystyle}\right]w^0 &
\left[\begin{displaystyle}-\left(3+\frac{1}{s^2}\right)\frac{(w^0)^2}{w^2}+1\end{displaystyle}\right]w^1 \\
\left[\begin{displaystyle}-\left(3+\frac{1}{s^2}\right)\frac{(w^0)^2}{w^2}+1\end{displaystyle}\right]w^1 &
\left[\begin{displaystyle}\left(3+\frac{1}{s^2}\right)\frac{(w^1)^2}{w^2}+1\end{displaystyle}\right]w^0
\end{pmatrix}
}
and
\eqn{A1def}{
 A^1 = \begin{pmatrix}
\left[\begin{displaystyle}\left(3+\frac{1}{s^2}\right)\frac{(w^0)^2}{w^2}-1\end{displaystyle}\right]w^1 &
-\left[\begin{displaystyle}\left(3+\frac{1}{s^2}\right)\frac{(w^1)^2}{w^2}+1\end{displaystyle}\right]w^0 \\
-\left[\begin{displaystyle}\left(3+\frac{1}{s^2}\right)\frac{(w^1)^2}{w^2}+1\end{displaystyle}\right]w^0 &
\left[\begin{displaystyle}\left(3+\frac{1}{s^2}\right)\frac{(w^1)^2}{w^2}+3\end{displaystyle}\right]w^1
\end{pmatrix},
}
where
\eqn{w2def}{
w^2 = (w^0)^2-(w^1)^2
}
and (see \eqref{fwexA5})
\eqn{sexpdef}{
s^2 = \frac{\gamma+1}{\gamma}\left(\begin{displaystyle}\frac{1}{(w^2)^{\frac{1}{2(\gamma+1)}}}-1\end{displaystyle}\right),
}
we can also write the the EFW equations \eqref{eul1} explicitly as
\eqn{EFWexp}{
A^0\del{0}\begin{pmatrix} w^0 \\ w^1 \end{pmatrix}
+ A^1\del{1}\begin{pmatrix} w^0 \\ w^1 \end{pmatrix} =0 .
}

In order to derive a formulation of the Euler equations that is
suitable to analyze the limit $s^2\searrow 0$, we use a particular frame formulation
of the EFW equations. First, we
set
\leqn{e0def}{
e_0 = e^\mu_0\del{\mu} = w = w^\mu \del{\mu},
}
where
\eqn{deldef}{
\del{\mu} = \frac{\partial \;}{\partial x^\mu},
}
and choose
\eqn{eJdef}{
e_1 = e_1^\mu \del{\mu}
} orthogonal to $e_0$ so that the frame metric
\eqn{fmetric1}{
g_{ij} =  g(e_i,e_j) = g_{\mu\nu} e^\mu_i e^\nu_j
}
satisfies
\leqn{fmetric2}{
g_{01}=g_{01} = 0,
}
and
\leqn{fmetric5}{
g_{00} = g(e_0,e_0) = w^2.
}
At the moment, we leave the length of $e_1$ unspecified. The freedom to fix the length of $e_1$ will be used below. From,
\eqref{gcart} and \eqref{wcartB}, it is clear that
\leqn{ecart}{
(e_1^\mu) = q(x^0,x^1)(v(x^0,x^1),1)
}
where $q(x^0,x^1)$ is a function to be determined.

%Writing the frame metric $g_{ij}$ in matrix form
%\eqn{fmetric3}{
%(g_{ij}) = \begin{pmatrix} g_{00} & 0 \\
%0 &  g_{11} \end{pmatrix}
%}
%the inverse frame metric $g^{ij}$ is given by
%\leqn{fmetric4}{
%(g^{ij}) = (g_{ij})^{-1} = \begin{pmatrix} \displaystyle{\frac{1}{g_{00}}} & 0\\
%0 &  \displaystyle{\frac{1}{g_{11}}} \end{pmatrix}
%}
%where we note that
%\leqn{fmetric5}{
%g_{00} = g(e_0,e_0) = w^2
%}by \eqref{e0def}.

Next, we denote the coframe by
\eqn{theta1}{
\theta^i = \theta^i_\mu dx^\mu \qquad (\theta^i_\mu e^\mu_j = \delta^i_j),
}
and let $\omega_i{}^k{}_j$ denote the connection coefficients so that
\eqn{omegadef1}{
\nabla_{e_i} e_j = \omega_i{}^k{}_j e_k.
}
We also define the connection 1-forms $\omega^k{}_j$ in the standard fashion
\eqn{omegadef2}{
\omega^k{}_j = \omega_i{}^k{}_j\theta^i,
}
and set
\eqn{omegadef3}{
\omega_{kj} = g_{kl}\omega^l{}_j = \omega_{ikj}\theta^i,
}
where
\leqn{omegadef4}{
\omega_{ikj} = g_{kl}\omega_i{}^l{}_j=\omega_{ikj} = g(\nabla_{e_i} e_j, e_k).
}

%Using
%\eqn{covder}{
%\nabla_{e_i} e_j = e_i^\mu \nabla_\mu e_j^\nu \del{\nu},
%}
%and \eqref{e0def}, we get from \eqref{omegadef4} that
%\leqn{omegadef5}{
%\omega_{kj0} = g( e_k^\mu \nabla_\mu w^\nu \del{\nu}, e_j^\gamma \del{\gamma}) = e_k^\mu \nabla_\mu w^\nu %e_j^\gamma g(\del{\nu},\del{\gamma}) = e_k^\mu \nabla_\mu w^\nu e_j^\gamma g_{\nu\gamma}.
%}

%Recalling the identities
%\lgath{fbrel}{
%e^\mu_i \theta_\mu^j = \delta^i_j, \quad e^\mu_i \theta^i_\nu = \delta^\mu_\nu, \quad %g^{\mu\lambda}g_{\lambda\nu}=\delta^\mu_\nu, \quad g^{ik}g_{kj} = \delta^i_j \label{fbrel.1} \\
%g_{ij} = e_i^\mu e_j^\nu g_{\mu\nu}, \quad \quad g_{\mu\nu} = \theta^i_\mu \theta^i_\nu g_{ij}, \quad g^{ij} = %\theta^i_\mu \theta^j_\nu g^{\mu\nu},
%\quad g^{\mu\nu} = e^\mu_i e_i^\nu g^{ij}, \label{fbrel.2}
%}
%we observe that
%\leqn{Adef0a}{
%\theta^k_\gamma w^\gamma = \theta^k_\gamma e^\gamma_0 = \delta^k_0,
%}
%and
%\leqn{Adef0b}{
%g^{ip}e_p^\mu w_\mu = g^{ip} e_p^\mu g_{\mu\nu} w^\nu = g^{ip} e^\mu_p \theta^k_\mu \theta^l_\nu g_{kl} w^\nu = %g^{ip} \delta_p^k g_{k0} = \delta^i_0.
%}

Letting
\eqn{Adef}{
A^{ijk} = g^{ip}g^{jq}e^\mu_p e^\nu_q \theta^k_\gamma A_{\mu \nu}{}^\gamma,
}
a short calculation using \eqref{eul3}, \eqref{e0def}, \eqref{fmetric2}, and \eqref{fmetric5} shows
that
\leqn{Adef2}{
A^{ijk}
%&=  \left(3+\frac{1}{s^2}\right) \frac{g^{ip}e_p^\mu w_\mu g^{jq}e_q^\nu w_\nu \theta^k_\gamma w^\gamma}{w^2}
%-g^{ip}g^{iq} e^\mu_p e^\nu_q \theta^k_\gamma \delta^\gamma_\nu w_\mu  \notag \\
%& \qquad - g^{ip}g^{iq}e^\mu_p e^\nu_q \theta^k_\gamma \delta^\gamma_\mu w_\nu - g^{ip}g^{iq}e^\mu_p e^\nu_q %\theta^k_\gamma w^\gamma g_{\mu\nu} \notag \\
%&
= \left(3+\frac{1}{s^2}\right)\frac{\delta^i_0\delta^j_0\delta^k_0}{g_{00}} - \delta^i_0 g^{jk} - g^{ik}\delta^j_0
-g^{ij}\delta^k_0 .
%\label{Adef2.1}.
}
Moreover, it follows from \eqref{e0def} and \eqref{omegadef4} that
\leqn{omegadef5}{
\omega_{kj0} = e_k^\mu \nabla_\mu w^\nu e_j^\gamma g_{\nu\gamma}.
}

%by \eqref{fmetric5} and \eqref{fbrel.1}-\eqref{Adef0b},
%and that
%\leqn{Adef3}{
%A^{ijk}\omega_{kj0} = g^{ip}g^{jq}e^\mu_p e^\nu_q \theta^k_\gamma A_{\mu \nu}{}^\gamma e_k^\lambda \nabla_\lambda %w^\delta e_j^\sigma g_{\delta\sigma}
% = g^{ip}e^\mu_p A_{\mu \delta}{}^{\lambda} \nabla_{\lambda} w^\delta
%}
%by \eqref{omegadef5}-\eqref{fbrel.2}.
Together, \eqref{Adef2}, \eqref{omegadef5} and the invertibility of $g^{ip}$ and $e^\mu_p$ show that the EFW equations \eqref{eul1} are equivalent to
\leqn{feul1}{
A^{ijk}\omega_{kj0} = 0.
}
Using \eqref{fwexA5}, \eqref{fmetric2}, \eqref{fmetric5}, and \eqref{Adef2}, the $i=0$ and $i=1$ components of \eqref{feul1}
are
\leqn{feul2}{
\omega_{000} - \frac{s^2 g_{00}}{g_{11}}\omega_{110} = 0 \AND
\omega_{100}+\omega_{010}  = 0,
}
respectively, where
\leqn{feul3}{
s^2 = \frac{\gamma+1}{\gamma}\left(\frac{1}{{g_{00}}^{\frac{1}{2(\gamma+1)}}}-1\right).
}
Also, due to the connection $\omega^i{}_j$ being metric, the frame metric satisfies (see \cite[Ch.V,\S B]{Choq})
\eqn{dg1a}{
\ed g_{jk} = \omega_{jk} + \omega_{kj}
}
or equivalently
\leqn{dg1}{
e_{i}(g_{jk}) = \omega_{ijk} + \omega_{ikj},
}
and
it follows immediately from \eqref{fmetric2} that
\leqn{framefix1}{
\omega_{001} + \omega_{010} = 0.
}

Since the connection $\omega^{k}{}_j$ is torsion free\footnote{This follows by virtue of the connection $\omega^i{}_j$ being metric.}, it satisfies the following
Cartan structure equation (see \cite[Ch.V,\S B]{Choq})
\eqn{cframe1a}{
\ed \theta^i + \omega^i{}_j\Wp \theta^j = 0,
}
or equivalently
\eqn{cframe1}{
[e_0,e_1] = \bigl(\omega_0{}^k{}_1 - \omega_1{}^k{}_0\bigr)e_k.
}
We rewrite this as follows:
\lalign{cframe2}{
[e_0,e_1] =  & -\frac{1}{g_{00}}\bigl(\omega_{010} + \omega_{100}\bigr)e_0+ \frac{1}{g_{11}}\bigl(\omega_{011} - \omega_{110}\bigr)e_1 &&\text{by \eqref{fmetric2} and \eqref{framefix1}} \notag\\
& = \frac{1}{g_{11}}\left(\omega_{011} - \frac{g_{11}}{s^2 g_{00}}\omega_{000}\right)e_1 && \text{by \eqref{feul2}} \notag\\
& =  \frac{1}{2g_{11}}\left( e_0(g_{11}) - \frac{g_{11}}{s^2 g_{00}}e_{0}(g_{00})\right)e_1. && \text{by \eqref{dg1}}. \label{cframe2.1}
}
Defining a function by
\eqn{cframe3}{
F(\xi) = \frac{1}{\xi^\frac{\gamma}{2(\gamma+1)} \left(\frac{1}{{\xi}^{\frac{1}{2(\gamma+1)}}}-1\right)^{\gamma}},
}
a short calculation shows that $F(\xi)$ satisfies
\eqn{cframe4}{
F'(\xi) = \frac{F(\xi)}{2\xi \frac{\gamma+1}{\gamma}\left(\frac{1}{{\xi}^{\frac{1}{2(\gamma+1)}}}-1\right)},
}
and hence
\eqn{cframe5}{
F'(g_{00}) = \frac{F(g_{00})}{2g_{00} s^2}
}
by \eqref{feul3}. But this implies that
\eqn{cframe6}{
e_0\bigl(\ln(F(g_{00})^2\bigr) = \frac{1}{s^2 g_{00}}e_{0}(g_{00}),
}
which we can in turn use to write \eqref{cframe2.1} as
\leqn{cframe7}{
[e_0,e_1] = \Half e_0\left(\ln\left(\displaystyle{\frac{-g_{11}}{F(g_{00})^2}}\right)\right)e_1 .
}
We now use the freedom to fix the length of $e_1$ by setting
\leqn{e1fix}{
g_{11} = -F(g_{00})^2.
}
By \eqref{cframe7}, we arrive at the equivalence of the EFW equations with the
vanishing of the following Lie bracket
\leqn{cframe8}{
[e_0,e_1] = 0.
}

We also note that if we set
\leqn{cframe10}{
u^{2\gamma} = \frac{1}{F(g_{00})^2},
}
then \eqref{cframe10} can be solved for $g_{00}$ to give
\leqn{cframe11}{
w^2 = g_{00} = (1-u)^{2(\gamma+1)},
}
which allows us, using \eqref{fwexA5}, to write the square of the sound speed as
\leqn{cframe12}{
s^2 = \frac{\gamma+1}{\gamma}\frac{u}{1-u}.
}
Using this and the normalization \eqref{e1fix} condition, we see
from \eqref{ecart} that
\eqn{ecartB}{
q =  \frac{1}{\sqrt{1-v^2} u^\gamma} = \frac{1}{\sqrt{1-v^2}
\left(\begin{displaystyle}\frac{s^2}{\frac{\gamma+1}{\gamma}+s^2}\end{displaystyle}\right)^\gamma},
}
and hence, that
\eqn{ecartC}{
(e_1^\mu) =
\frac{(1+K\rho(x^0,x^1)^\frac{1}{\gamma})^\gamma}{K^\gamma \rho(x^0,x^1) \sqrt{1-v(x^0,x^1)^2}}
\bigl(v(x^0,x^1),1\bigr)
}
by \eqref{sound}.

To summarize, the main result of this section is that the EFW equations
are equivalent to the vanishing of the Lie bracket \eqref{cframe8} which in components reads
\eqn{cframe8a}{
e^\mu_0 \del{\mu} e^\nu_1 - e^\mu_1 \del{\mu} e^\nu_0 = 0,
}
where the frame components $e^\mu_i$ are given by
\lalign{fcart}{
(e_0^\mu) & = \frac{1}{\bigl(1+K\rho(x^0,x^1)^{\frac{1}{\gamma}}\bigr)^{\gamma+1}\sqrt{1-v(x^0,x^1)^2}}\bigl( 1 , v(x^0,x^1)\bigr)
\label{fcart.1},\\
(e_1^\mu) &=
\frac{(1+K\rho(x^0,x^1)^\frac{1}{\gamma})^\gamma}{K^\gamma \rho(x^0,x^1) \sqrt{1-v(x^0,x^1)^2}}
\bigl(v(x^0,x^1),1\bigr). \label{fcart.2}
} 

\sect{sym}{Existence of solutions to the Euler equations}

\subsect{Ewave}{A wave equation formulation of the EFW equations}

The vanishing of the Lie bracket \eqref{cframe8} implies the existence of coordinates which, at least locally,
trivialize the vector fields $e_0$ and $e_1$. To construct these coordinates, we let
\eqn{FlA}{
\Fc_{i,\tau}(\xh^0,\xh^1) = \bigl(\Fc^0_{i,\tau}(\xh^0,\xh^1),\Fc^1_{i,\tau}(\xh^0,\xh^1)\bigr)
\qquad i=0,1
}
denote the flow maps of the vector fields $e_i$ $(i=0,1)$, that is  $\Fc_{i,\tau}$ is the unique solution
to the initial value problem
\alin{FlB}{
\frac{d\;}{d\tau} \Fc^\mu_{i,\tau}(\xh^0,\xh^1) & = e^\mu_i\bigl(\Fc_{i,\tau}(\xh^0,\xh^1)\bigr)
\qquad \mu=0,1, \\
 \Fc_{i,0}(\xh^0,\xh^1) &= (\xh^0,\xh^1),
}
where the $e^\mu_i$ are given explicitly by the formulas \eqref{fcart.1}-\eqref{fcart.2}.
By translating, if necessary, we can assume the origin $(x^0,x^1)=(0,0)$ is in the domain on which
the vector fields $e^\mu_i$ are defined. We then introduce a change of coordinates by the formula
\leqn{hcoord1}{
(x^0,x^1) = \Psi(\xh^0,\xh^1) = \Fc_{0,\sqrt{\frac{\gamma}{\gamma+1}}\xh^0}\circ \Fc_{1,\xh^1-c}(0,0).
}
where $c$ is a constant\footnote{The constant $c$ is chosen so that the point $\lim_{\xh^1\searrow 0} \Fc_{1,\xh^1-c}(0,0)$
lies on the vacuum boundary where $\rho = 0$.}.

Since the vector fields $w=e_0$, and $e_1$ commute, if we
define
\eqn{hcoord2}{
\wh = \eh_0 = \Psi^* e_0 \AND \eh_1 = \Psi^* e_1,
}
then
\leqn{coord1}{
\wh = \eh_0 =  \sqrt{\frac{\gamma+1}{\gamma}}\delh{0} \AND \eh_1 = \delh{1},
}
where
\eqn{hcoord3}{
\delh{\mu} =\frac{\partial \;}{\partial \xh^\mu}.
}
As we show below (see Remark \ref{vbrem}), the vacuum boundary where $\rho$ vanishes is contained in the set $\xh^1 =0$. Since
the two-velocity $\vh^\mu$ is given by $\vh^\mu = (\wh)^{-1}\wh^\mu$, equation \eqref{coord1} shows
that the vacuum boundary moves with the fluid as expected.

Next, defining
\eqn{guhatdef}{
\gh = \Psi^*g = \gh_{\mu\nu} d\xh^\mu d\xh^\nu \AND
\uh = \Psi^* u,
}
it follows directly from  \eqref{coord1},
\eqref{e1fix}, \eqref{cframe10}, and \eqref{cframe11} that\footnote{The metric $\gh = \Psi^*g$ is just the original Minkowski metric $g$
(see \eqref{gcart}) expressed in the $(\xh^\mu)$  coordinates defined by \eqref{hcoord1}. }
\leqn{met}{
\gh= \frac{\gamma(1-\uh)^{2(\gamma+1)}}{\gamma+1}d \xh^0 d \xh^0 - \frac{1}{\uh^{2\gamma}} d \xh^1 d\xh^1.
}
Computing the Ricci scalar of this metric, we find that
\leqn{Ric}{
\Rh = \frac{\uh^{\gamma}}{G(\uh)^{1/2}}\left[2(\gamma+1)\delh{0}\left(\frac{1}{\uh^{\gamma+1}G(\uh)^{1/2}}\delh{0}\uh\right)
+ \delh{1}\left(\frac{\uh^\gamma G'(\uh)}{G(\uh)^{1/2}}\delh{1}\uh\right)\right],
}
where
\eqn{coord4}{
G(\uh) = (1-\uh)^{2(\gamma+1)}.
}

In order to simplify the following calculation, we will from this point on assume that $\gamma =1$, and define
\leqn{z1}{
\zh = \arcsin(-1+2\uh)+\frac{\pi}{2},
}
which can be inverted to give
\leqn{z2}{
\uh = \Half(1-\cos(\zh)).
}

\begin{rem}
To extend the analysis to $\gamma > 0$, the appropriate $\zh$ variable that replaces \eqref{z1} can be obtained by
solving the initial value problem
\eqn{zdiff1}{
\frac{d\zh}{d\uh} = \left(-\frac{G'(\uh)}{2(\gamma+1)\uh G(\uh)} \right)^{1/2} \quad : \quad \zh(0) = 0
}
for $\uh\geq 0$. A solution to this initial value problem yields the identity
\eqn{zdiff2}{
\frac{1}{\uh^{\gamma+1}G(\uh)^{1/2}}\frac{d\zh}{d\uh} = \left(-\frac{\uh^\gamma G'(\uh)}{2(\gamma+1)G(\uh)^{1/2}}
\frac{d\zh}{d\uh}\right)^{-1},
}
which allows a similar analysis as in the $\gamma =1$ case to be used.
\end{rem}

In terms of the $\zh$ variable, the metric \eqref{met} and the Ricci scalar \eqref{Ric} become
\leqn{zmet}{
\gh= \frac{(1+\cos(\zh))^{4}}{32}d \xh^0 d \xh^0 - \frac{4}{(1-\cos(\zh))^{2}} d \xh^1 d \xh^1,
}
and
\leqn{zRic}{
\Rh = \frac{(1-\cos(\zh))}{2(1+\cos(\zh))^2}\left[\delh{0}\left(\frac{8}{\sin(\zh)^3}\delh{0}\zh\right) - \delh{1}\left(\frac{\sin(\zh)^3}{8}\delh{1}\zh\right)\right],
}
respectively. But the metric \eqref{zmet} is flat\footnote{We recall that the Ricci scalar is a coordinate invariant which
can be stated as $\Rh = R\circ \Psi$. Since
$g$ is the Minkowski metric and consequently flat, its Ricci scalar vanishes. By the coordinate invariance,
the Ricci scalar of $\gh$ must also vanish. }, and so \eqref{zRic} and the above arguments show that the EFW equations \eqref{eul3}
are equivalent to the wave equation
\leqn{zwave1}{
\delh{0}\left(\frac{8}{\sin(\zh)^3}\delh{0}\zh\right) - \delh{1}\left(\frac{\sin(\zh)^3}{8}\delh{1}\zh\right) = 0.
}
\begin{rem}\label{zwave1rem}
It is well known that in $1+1$ dimensions\footnote{In fact, this is true for irrotational fluids in any dimension.} the Euler equations can be reduced to a quasi-linear scalar wave equations of the form (see
\cite{Christ} or \cite{VMP} for details)
\eqn{irrwave1}{
\nabla_\mu \bigl(H(|\nabla\varphi|^2)\nabla^\mu\varphi) =0  \qquad (|\nabla\varphi|^2 = g^{\mu\nu}\del{\mu}\varphi\del{\nu}\varphi),
}
where $H(\cdot)$ is a particular function determined by the equation of state, and the proper energy density $\rho$ and two-velocity $v^\mu$ can be recovered via the formulas
\eqn{irrwave2}{
|\nabla\varphi|^2 = \exp\left(\int_0^p \frac{dp}{\rho(p)+p}\right) \AND v^\mu = \frac{1}{|\nabla\varphi |} g^{\mu\nu}\del{\mu} \varphi.
}
In light of this, it is perhaps not surprising that we are able to reduce the Euler equations
to a quasi-linear scalar wave equation. What is new here is that proper energy density
is a function of the scalar $\zh$, and it vanishes where $\zh$ vanishes. As will be
shown below, this makes the wave equation \eqref{zwave1} particularly well suited for analyzing the vacuum boundary problem.
\end{rem}

Defining an auxiliary Lorentzian metric $\lambdah$ by\footnote{The metric $\lambdah_{\mu\nu}$ is conformal to the acoustical metric $\hat{\alpha}_{\mu\nu}
= \hat{s}^2 \hat{v}_\mu \hat{v}_\nu + \hat{h}_{\mu\nu}$ with the relation between the two given by $\hat{\alpha}_{\mu\nu} = \frac{\sin^3(\zh)}{2(1-\cos(\zh))}\lambdah_{\mu\nu}$.}
\leqn{lamdef}{
\lambdah = \lambdah_{\mu\nu}d \xh^\mu d \xh^\mu = \frac{\sin(\zh)^3}{8} d \xh^0 d \xh^0 -
\frac{8}{\sin(\zh)^3} d \xh^1 d \xh^1,
}
equation \eqref{zwave1} is easily seen to be equivalent to
\leqn{zwave2}{
\Box_{\lambdah} \zh = \frac{1}{\sqrt{-\lambdah}}\delh{\mu}\left(\sqrt{-\lambdah}\lambdah^{\mu\nu} \delh{\nu} \zh\right) = 0,
}
where
\eqn{sqrtlamdef}{
\sqrt{-\lambdah} = \sqrt{-\det(\lambdah_{\mu\nu})} \AND \lambdah^{\mu\nu} = (\lambdah_{\mu\nu})^{-1}.
}

\subsect{null}{Conformal coordinates}

We introduce a second change of coordinates
\leqn{cbar1}{
(\xh^0,\xh^1) = \Psib(\xb^0,\xb^1) = \left(\psib(\xb^0,\xb^1),\phib(\xb^0,\xb^1)\right)
}
and let
\leqn{cbar2}{
\zb = \Psib^*\zh.
}
Pulling back the metric \eqref{lamdef} using the diffeomorphism \eqref{cbar1}, we find, using \eqref{cbar2}, that
\leqn{lambd1}{
\lambdab = \Phi^*\lambdah = \lambdab_{\mu\nu} d \xb^\mu d \xb^\nu,
}
where
\eqn{lambd2}{
(\lambdab_{\mu\nu}) = \begin{pmatrix}\begin{displaystyle} \mu(\delb{0}\psib)^2
- \frac{1}{\mu}(\delb{0}\phib)^2  \end{displaystyle} & \begin{displaystyle}  \mu\delb{0}\psib\delb{1}\psib
- \frac{1}{\mu}\delb{0}\phib\delb{1}\phib  \end{displaystyle} \\
\begin{displaystyle}  \mu\delb{0}\psib\delb{1}\psib
- \frac{1}{\mu}\delb{0}\phib\delb{1}\phib  \end{displaystyle} &
\begin{displaystyle}  \mu(\delb{1}\psib)^2
- \frac{1}{\mu}(\delb{1}\phib)^2  \end{displaystyle}
\end{pmatrix},
}
\eqn{mudef}{
\mub = \frac{\sin^3(\zb)}{8},
}
and
\eqn{delbdef}{
\delb{\mu} = \frac{\partial \;}{\partial \xb^\mu}.
}

We fix the choice of coordinates by requiring that the metric \eqref{lambd1} is conformally flat in these
coordinates. We accomplish this by demanding that $\phib$ and $\psib$ satisfy
\leqn{cfix}{
\mub \delb{0}\psib = \delb{1}\phib \AND  \delb{0}\phib = \mub \delb{1}\psib.
}
With this choice, the  metric \eqref{lambd1} can be written as
\leqn{lambd3}{
\lambdab = \Omegab \mb,
}
where
\leqn{mdef}{
\mb = d \xb^0 d \xb^0 - d \xb^1 d \xb^0,
}
and
\eqn{Omegadef}{
\Omegab = \frac{\bigl((\del{1}\phib)^2-(\del{0}\phib)^2\bigr)}{\mub} .
}
Also, from the Jacobian of the transformation \eqref{cbar1}
\leqn{Jdef}{
J_{\Psib} = \begin{pmatrix} \begin{displaystyle}\delb{0}\psib \end{displaystyle}
& \begin{displaystyle} \delb{1}\psib \end{displaystyle}\\
\delb{0}\phib & \delb{1}\phib \end{pmatrix},
}
we see that
\leqn{detJ}{
\det J_{\Psib} = \Omegab.
}

\subsect{cwave}{The conformal wave equation}

By the conformal invariance of the wave equation in 2 dimensions\footnote{The conformal invariance follows directly from
the identity $\Box_{\Omegab \mb}\zb = \frac{1}{\Omegab}\Box_{\mb}\zb$. See Appendix D of \cite{Wald} for a derivation of this identity.}, and the invertibility of the coordinate transformation
\eqref{cbar1} on regions where $\det\! J_{\Psib}$ does not vanish, it follows from \eqref{lambd3} and \eqref{detJ} that
the wave equation \eqref{zwave2} is equivalent to
\leqn{zwave3}{
\Box_{\mb} \zb = \delb{0}^2 \zb - \delb{1}^2\zb = 0.
}
We will solve this system on the spacetime region
\eqn{Udef}{
U = \{\, (\xb^0,\xb^1) \, | \, \xb^0,\xb^1 > 0 \}
}
with the boundary condition
\eqn{bczwave}{
\zb|_{\Gamma} = 0,
}
where
\eqn{Sigmadef}{
\Gamma = \{ \, (\xb^0,0)\, |\, \xb^0 > 0\}.
}
Since the vanishing of $\zb$ implies the vanishing of the proper energy density $\rhob$ (see \eqref{sound},
\eqref{cframe12}, \eqref{z1}, and \eqref{z2}),
the boundary condition will ensure the the vacuum boundary lies in $\Gamma$.

We let
\eqn{SigmadefA}{
\Sigma = \{ \, (0,\xb^1) \, |\, \xb^1 > 0\}
}
denote the initial hypersurface, and
and we define the following related sets
\eqn{UTdef}{
U_{T,\delta} = \{\, (\xb^0,\xb^1) \, | \, 0<\xb^0 < T \AND 0<\xb^1<\delta \,\},
}
\eqn{GammaTdef}{
\Gamma_{T} = \{\, (\xb^0,0) \, | \, 0<\xb^0 < T \,\},
}
and
\eqn{Sigmaddef}{
\Sigma_{\delta} = \{\, (0,\xb^1) \, | \, 0<\xb^1<\delta \,\}.
}

\begin{prop}\label{eprop}
Suppose $k\in \Nbb$, $s>1/2+k$, $\zb_0 \in H^{s}(\Sigma)$, $\zb_1\in H^{s+1}(\Sigma)$, and the
\eqn{eprop1}{
\zb_{\ell} = \del{\xb^1}^2\zb_{\ell-2} \quad \ell = 2,3,\ldots, s
}
satisfy the compatibility conditions
\eqn{eprop0}{
\zb_\ell |_{\xb^1=0} = 0 \quad \ell = 0,1,\ldots, s
}
where $\Sigma$ and $\Gamma$ meet.
Then
\begin{itemize}
\item[(i)] there exists
a unique solution $\zb \in C^k(U)$ to the initial boundary value problem
\lalign{eprop3}{
\Box_{\mb} \zb & = 0,  \label{eprop3.1} \\
\zb\bigl|_{\Gamma} & = 0 \label{eprop3.2},\\
\zb\bigl|_{\Sigma} & = \zb_0, \label{eprop3.3}\\
\delb{0}\zb \bigl|_{\Sigma} & = \zb_1, \label{eprop3.4}
}
\item[(ii)]
for any $T,\delta>0$,
\eqn{eprop4}{
\zb \in C^k(U_{T,\delta})\cap W^{k,\infty}(U_{T,\delta}),
}
 and
\item[(iii)] if $0<2\kappa \leq \delb{1}\zb|_{\Sigma_\delta} \leq  \beta$ and $|\delb{0}\zb|_{\Sigma_\delta} < \kappa/4$ for some $\kappa,\delta,\beta >0$ , then there exists a $T_0$ such that
\gath{eprop5}{
\bigl((\delb{1}\zb)^2-(\delb{0}\zb)^2\bigr)\bigl|_{U_{T_0,\delta}} \geq \kappa^2/4 > 0, \\
0<\kappa \leq \delb{1}\zb\bigl|_{U_{T_0,\delta}} \leq 2\beta,
\intertext{and}
\kappa \xb^1 \leq \zb(\xb^0,\xb^1) \leq 2\beta \xb^1
}
for all $(\xb^0,\xb^1)\in U_{T_0,\delta}$.
\end{itemize}
\end{prop}
\begin{proof}
Statements (i) and (ii) follows from standard linear hyperbolic theory for initial boundary value problems. For example, see Theorems 3, 4, 5 and 6 in Section 7.2 of \cite{Evans}. The first two bounds from statement (iii) follow from the bounds on the initial data and the continuous
differentiability of $\zb(\xb^0,\xb^1)$. The last bound in (iii) follows from integrating the second with respect to $\xb^1$ while using the
fact that $\zb(\xb^0,0)=0$.
\end{proof}

\begin{rem} \label{eremA}$\;$
\begin{itemize}
\item[(i)] Clearly, we can explicitly solve the initial boundary value problem \eqref{eprop3.1}-\eqref{eprop3.2} using the well known formula
\leqn{eremA.1}{
\zb(\xb^0,\xb^1) = \begin{cases} \Half\left( \zb_0(\xb^0+\xb^1)-\zb_0(\xb^0-\xb^1) +
\int_{\xb^0-\xb^1}^{\xb^0+\xb^1} \zb_1(\xi)\, d\xi\right) \quad \xb^0 < \xb^1, \; 0<\xb^1 \\
\Half\left( \zb_0(\xb^0+\xb^1)+\zb_0(\xb^1-\xb^0) + \int_{\xb^1-\xb^0}^{\xb^0+\xb^1} \zb_1(\xi)\, d\xi\right) \quad \quad \xb^1<\xb^0 , \; 0<\xb^1  \end{cases}.
}
The point of Proposition \ref{eprop} is that it extends in the obvious manner if we replace $U$ by an open set that
is bounded by a timelike hypersurface $\Gamma$ and a spacelike hypersurface $\Sigma$. In this situation, there is no equivalent to
the simple formula \eqref{eremA.1} for solutions to the initial boundary value problem \eqref{eprop3.1}-\eqref{eprop3.2}. As we shall see below, the only  property
of $U$ that we need is
\eqn{eremA.2}{
\{\: \xi\xb = (\xi\xb^0,\xi\xb^1) \; |  \; \xb \in U \AND 0<\xi \leq 1 \:\} \subset U.
}
In fact, this can be weakened to the existence of a $\epsilon >0$ small enough so that
\eqn{eremA.3}{
\{\: \xi\xb = (\xi\xb^0,\xi\xb^1) \; |  \; \xb\in U\cap B_{\epsilon}(0) \AND 0<\xi<\epsilon \:\} \subset U,
}
where $B_\epsilon(0) = \{ \: \xb\in \Rbb^2 \;|\; |\xb|= \sqrt{(\xb^0)^2+(\xb^1)^2} <\epsilon \:\}$.
\item[(ii)] It is not difficult to show that the conditions imposed on the initial data in Proposition \ref{eprop} are satisfied
for a wide class of initial data. For example, initial data of the form
\eqn{eremA.4}{
\zb_0 = cy+ p(y), \quad \zb_1 = q(y) \quad (c=\text{const}>0),
}
where
\begin{itemize}
\item[(ii.a)]
\eqn{eremA.5}{
p(y) = \text{o}(y^{s}) \AND q(y) = \text{o}(y^{s-2}) \quad \text{as $y\searrow 0$}
}
if $s$ is an even integer, and
\item[(ii.b)]
\eqn{eremA.6}{
p(y) = \text{o}(y^{s-1}) \AND q(y) = \text{o}(y^{s-1}) \quad \text{as $y\searrow 0$}
}
if $s$ is an odd integer,
\end{itemize}
satisfy all the conditions on the initial data in Proposition \ref{eprop}.
\end{itemize}
\end{rem}

\subsect{ccord}{Fixing the conformal coordinates}
Letting $*_{\mb}$ denote the Hodge dual operator\footnote{We recall that the Hodge dual operator on one forms
is defined by $*_{\mb}\ed \xb^\mu = \sqrt{|\mb|}\mb^{\mu\nu}\epsilon_{\nu\sigma} \ed \xb^{\sigma}$
where $|\mb|=-\det(\mb_{\mu\nu})$ and $\epsilon_{\nu\sigma}$ is the completely antisymmetric symbol. In
particular, this implies that $*_{\mb}\ed\xb^0=\ed\xb^1$ and $*_{\mb}\ed \xb^1 = \ed \xb^0$.} of the metric \eqref{mdef},
we can write \eqref{zwave3} as $*\ed(*\ed\zb)=0$ which  implies that the one form $*\ed\zb$ is closed.  Consequently, we
get by the proof of the Poincar\'{e} lemma (see Ch. V \S4, Theorem 4.1 of \cite{Lang}) that
\leqn{zdef}{
z(\xb) = \int_0^1 \delb{1}\zb(\xi\xb)\,\xb^0 + \delb{0}\zb(\xi\xb)\,\xb^1 \, d\xi + z_0  \qquad (z_0\in\Rbb)
}
satisfies
\leqn{dz}{
\ed z = *\ed\zb,
}
or equivalently
\leqn{dz1}{
\delb{0}z = \delb{1}\zb \AND \delb{1}z = \delb{0}\zb.
}
Evaluating $z$ at $x^0=0$, we see that
\eqn{z=0A}{
z(0,\xb^1) = \int_0^1 \zb_1(0,\xi\xb^1) \, \xb^1 \, d\xi + z_0.
}
Since $\zb_1 = \text{o}(\xb^1)$ as $\xb^1\searrow 0$ (see Remark \ref{eremA}.(ii)), it is clear that by choosing $z_0$ to be
\eqn{z0def}{
z_0 = -\min\left\{0, \inf_{0<\xb^1<\delta} \int_0^1 \zb_1(0,\xi\xb^1) \, \xb^1 \, d\xi \right\}
}
we can ensure that
\eqn{z=0B}{
0\leq z(0,\xb^1) \leq \delta \quad \text{for $0<\xb^1 < \delta$}.
}
It then follows from the integrating the first equation of \eqref{dz1} and the bounds on  $\del{1}\zb$
from Proposition \ref{eprop}.(iii) that
\leqn{z=0C}{
\kappa \xb^0 \leq z(\xb^0,\xb^1) \leq 2\beta \xb^0 + \delta
}
for all $(\xb^0,\xb^1) \in U_{T_0,\delta_0}$.

Since $\ed z \Wp \ed \zb = \bigl((\delb{1}\zb)^2-(\delb{0}\zb)^2\bigr)\ed \xb^0\Wp \ed \xb^1$,
the non-vanishing of $\bigl((\delb{1}\zb)^2-(\delb{0}\zb)^2\bigr)$ on $U_{T_0,\delta}$ (see Proposition \ref{eprop})
shows that $\{\ed z , \ed \zb\}$ forms a basis for the space of one forms at every point of $U_{T_0,\delta}$. This
allows us to look for solutions to \eqref{cfix} that are of the form
\leqn{zc3}{
\phib = \phi(z,\zb) \AND \psib = \psi(z,\zb).
}
To see this, we get from \eqref{dz} and \eqref{zc3} that
\leqn{zc4}{
*_{\mb} \ed \phib = \del{z}\phi \ed \zb + \del{\zb}\phi \ed z,
}
and
\leqn{zc5}{
\ed \psib = \del{z} \psi \ed z + \del{\zb} \psi\ed \zb.
}
Writing \eqref{cfix} as
\leqn{cfix1}{
\mu(\zb) \ed \psib = *_{\mb} \ed \phib,
}
where
\eqn{ubfunc}{
\mu(\zeta) =  \frac{\sin^3(\zeta)}{8},
}
we see from \eqref{zc4}, \eqref{zc5}, and \eqref{cfix1} that $\psi$ and $\phi$ satisfy
\leqn{cfix2}{
\mu(\zb) \del{z}\psi = \del{\zb}\phi \AND \mu(\zb) \del{\zb}\psi = \del{z} \phi.
}
From these equations, we then get the wave equation
\leqn{cfix3}{
\del{z}^2 \phi - \mu(\zb) \del{\zb}\left( \frac{1}{\mu(\zb)} \del{\zb} \phi \right) = 0
}
for $\phi$.

\begin{prop} \label{esprop}
Suppose the initial data $(\phi|_{z=0},\del{z}\phi|_{z=0})$ for the wave equation \eqref{cfix3} satisfies\footnote{The spaces
$\Hc^k$ are defined in Section \ref{wexist} of the Appendix.}
\lgath{exist1}{
\left(\left.\frac{\phi}{\sqrt{\mu(\zb)}}\right|_{z=0},\del{z}\left.\left(\frac{\phi}{\sqrt{\mu(\zb)}}\right)\right|_{z=0}\right) \in
\Hc^6\times \Hc^5,\label{exist1.1}\\
\frac{1}{\mu(\zb)}\del{\zb}\phi(0,\zb) \geq  c> 0, \label{exist1.2}
\intertext{and}
\del{z}\phi(0,\zb)  \geq 0 \label{exist1.3}
}
for all $0<\zb < \pi/2$. Then there exists a unique solution $\phi(z,\zb)$ to \eqref{cfix3} that can be written as
\leqn{exist2}{
\phi(z,\zb) = \phibr(z,\zb^4),
}
where $\phibr(t,\xi)$ satisfies
\leqn{exist3}{
\phibr \in C^1\bigl([0,\infty),C^{1,1/2}(0,(\pi/2)^4)\bigr)\cap \bigcap^5_{j=0} C^{j}\bigl([0,\infty),C^{5-j}(0,(\pi/2)^4)\bigr)
}
and
\leqn{exist4}{
|\del{t}\phibr(t,\xi)| + |\phibr(t,\xi)| \lesssim \xi.
}
Moreover, there exists a $\tau_0$ such that
\leqn{exist5}{
\del{\xi} \phibr(t,\xi) \geq c/2}
for all $(t,\xi)\in [0,\tau_0)\times (0,\pi/2)$.
\end{prop}
\begin{proof}
This is just a restatement of Theorem \ref{regthm} from the Section \ref{reg} of the Appendix.
\end{proof}

Integrating the first equation in \eqref{cfix2} with respect to $z$, and using the second
equation of \eqref{cfix2} to fix the undetermined function of integration, we find the following expression
\leqn{exist6a}{
\psi(z,\zb) = \int_{0}^z \frac{1}{\mu(\zb)}\del{\zb}\phi(\tau,\zb) d\tau +
\int_0^{\zb} \frac{1}{\mu(\zeta)}\del{z}\phi(0,\zeta) d\zeta
}
for $\psi$. Using \eqref{exist3}, we can write $\psi$ as
\leqn{exist6}{
\psi(z,\zb) = \frac{32\zb^3}{\sin^3(\zb)}\int_{0}^z \del{\xi}\phibr(\tau,\zb^4) d\tau +
\int_0^{\zb} \frac{8}{\sin^3(\zeta)}\del{t}\phibr(0,\zeta^4) d\zeta.
}

\begin{lem} \label{Jaclem} Suppose $k=3$, $\zb$ is the solution to the wave equation \eqref{zwave3} from Proposition \ref{eprop},
$z$ is a defined by \eqref{zdef}, $\phi$ and $\phibr$ are the maps from Proposition \ref{esprop}, and
$\psi$ is given by \eqref{exist6}. Then for $T_0$ and $\delta_0$ small enough, the change of coordinates map \eqref{cbar1}
\eqn{Jaclem1}{
\Psib\; : \; U_{T_0,\delta} \longrightarrow \Rbb^2 \: :\: (\xb)\longmapsto
(\psib(\xb),\phib(\xb)) = \bigl(\psi\bigl(z(\xb),\zb(\xb)\bigr),\phi\bigl(z(\xb),\zb(\xb)\bigr)\bigr)
}
is well defined and of class $C^3$ on $U_{T_0,\delta}$, and the Jacobian matrix is given by the formula
\eqn{Jaclem3a}{
J_{\Psib}(\xb) = \begin{pmatrix} \begin{displaystyle} \frac{\del{\zb}\phi(z(\xb),\zb(\xb))}{\mu(\zb(\xb))}
\end{displaystyle} & \begin{displaystyle} \frac{\del{z}\phi(z(\xb),\zb(\xb))}{\mu(\zb(\xb))} \end{displaystyle}\\
\del{z}\phi(z(\xb),\zb(\xb)) & \del{\zb}\phi(z(\xb),\zb(\xb)) \end{pmatrix} \begin{pmatrix}\delb{1}\zb(\xb) & \delb{0}\zb(\xb) \\
\delb{0}\zb(\xb) & \delb{1} \zb(\xb) \end{pmatrix},
}
and satisfies
\eqn{Jaclem2a}{
\det{J_{\Psib}}|_{U_{T_0,\delta}} > 0.
}
Furthermore,
\eqn{Jaclem3b}{
\Psib(\xb^0,0) = \left(32 \int_0^{\xb^0 \int_0^1\delb{1}\zb(\xi\xb^0,0)d\xi + z_0 } \del{\xi}\phibr(\tau,0) d\tau,0\right).
}
\end{lem}
\begin{proof}
By Proposition \ref{eprop}.(iii) and \eqref{z=0C}, we have that
\eqn{zzbran1}{
(z(\xb),\zb(\xb)) \in [0,2\beta T_0 + \delta)\times (0,2\beta\delta)
}
for all $\xb \in U_{T_0,\delta}$. Choosing $\delta$ and $T_0$ small enough, it
is clear that we can arrange that
\leqn{zzbran2}{
(z(\zb),\zb(\xb)) \in [0,\tau_0)\times (0,2\beta\delta) \subset [0,\tau_0)\times (0,\pi/2)
}
for all $\xb \in U_{T_0,\delta}$. As a consequence, the change of coordinates map (see \eqref{cbar1})
\leqn{Psibmap1}{
(\xh) = \Psib(\xb) = (\psib(\xb),\phib(\xb)) := \bigl(\psi\bigl(z(\xb),\zb(\xb)\bigr),\phi\bigl(z(\xb),\zb(\xb)\bigr)\bigr)
}
is well defined for all $\xb \in U_{T_0,\delta}$

A short calculation using \eqref{Jdef} and \eqref{Psibmap1} shows that
\eqn{Jaclem2}{
J_{\Psib} = \begin{pmatrix} \del{z}\psi(z,\zb) & \del{\zb}\psi(z,\zb)\\
\del{z}\phi(z,\zb) & \del{\zb}\phi(z,\zb) \end{pmatrix} \begin{pmatrix}\delb{0}z & \delb{1}z \\
\delb{0}\zb & \delb{1} \zb \end{pmatrix}.
}
Using \eqref{dz1} and \eqref{cfix2}, we can write this as
\leqn{Jaclem3}{
J_{\Psib} = \begin{pmatrix} \begin{displaystyle} \frac{\del{\zb}\phi(z,\zb)}{\mu(\zb)}
\end{displaystyle} & \begin{displaystyle} \frac{\del{z}\phi(z,\zb)}{\mu(\zb)} \end{displaystyle}\\
\del{z}\phi(z,\zb) & \del{\zb}\phi(z,\zb) \end{pmatrix} \begin{pmatrix}\delb{1}\zb & \delb{0}\zb \\
\delb{0}\zb & \delb{1} \zb \end{pmatrix}.
}
Taking the determinant gives
\eqn{Jaclem4}{
\det\bigl(J_{\Psib}\bigr) = \mu(\zb)\left( \left(\frac{\del{\zb}\phi(z,\zb)}{\mu(\zb)}\right)^2-
\left(\frac{\del{z}\phi(z,\zb)}{\mu(\zb)}\right)^2 \right)\bigl( (\delb{1}\zb)^2-(\delb{0}\zb)^2\bigr).
}

It follows from \eqref{exist2} and \eqref{exist5} that there exists a positive constant $C$ such that
\eqn{Jaclem5}{
\frac{\del{\zb}\phi(z,\zb)}{\mu(\zb)} = \frac{4\zb^3}{\mu(\zb)}\del{\xi}\phibr(z,\zb^4) \geq C > 0
}
for all $(z,\zb) \in [0,\tau_0)\times (0,\pi/2)$. This, in turn, implies via \eqref{zzbran2} that
\leqn{Jaclem6}{
\frac{\del{\zb}\phi(z(\xb),\zb(\xb))}{\mu(\zb(\xb))} = \frac{4\zb^3(\xb)}{\mu(\zb(\xb))}\del{\xi}\phibr(z(\xb),\zb^4(\xb)) \geq C > 0
}
for all $\xb \in  U_{T_0,\delta}$.

Next, we observe that
\eqn{Jaclem7}{
\left|\frac{\del{z}\phi(z,\zb)}{\mu(\zb)}\right| = \left|\frac{\del{t}\phibr(z,\zb^4)}{\mu(\zb)}\right| \lesssim
\frac{|\zb^4|}{\mu(\zb)}
}
for all $(z,\zb) \in [0,\tau_0)\times (0,\pi/2)$ by \eqref{exist2} and \eqref{exist4}. This inequality,  with the
help of \eqref{zzbran2} and the fact that $\lim_{\zb\searrow 0} \zb^4/\mu(\zb) = 0$, shows that by choosing
$\delta$ small enough, we can arrange that
\leqn{Jaclem8}{
\left|\frac{\del{z}\phi(z(\xb),\zb(\xb))}{\mu(\zb(\xb))}\right| \leq \frac{C}{\sqrt{2}}
}
for all $\xb \in U_{T_0,\delta}$.

The two inequalities \eqref{Jaclem6} and \eqref{Jaclem8} together with the bound
on $(\delb{1}\zb)^2-(\delb{0}\zb)^2$  from Proposition \ref{eprop}.(iii) and the formula \eqref{Jaclem3} show
that
\eqn{Jaclem9}{
J_{\Psib}|_{U_{T_0,\delta}} > 0.
}

From \eqref{exist2} and \eqref{exist4}, we see that
\eqn{Jaclem10}{
|\phi(z,\zb)| = |\phibr(z,\zb^4)| \lesssim |\zb^4|.
}
Since $\zb(\xb^0,0)=0$, it follows that
\leqn{Jaclem11}{
\phi(z(\xb^0,0),\zb(\xb^0,0)) = 0.
}

Using \eqref{exist4} again, we have that
\eqn{Jaclem12}{
\left| \int_0^{\zb} \frac{1}{\mu(\zeta)} \del{t}\phibr(0,\zeta^4)d\zeta\right|
\lesssim \int_0^{\zb} \zeta d\zeta \lesssim \zb^2.
}
This together with \eqref{zdef} and \eqref{exist6} shows that
\leqn{Jaclem13}{
\psi(z(\xb^0,0),\zb(\xb^0,0)) = 32 \int_0^{\xb^0 \int_0^1\delb{1}\zb(\xi\xb^0,0)d\xi + z_0 } \del{\xi}\phibr(t,0) d\tau.
}
From \eqref{Jaclem11} and \eqref{Jaclem13}, we see that
\eqn{Jaclem14}{
\Psib(\xb^0,0) = \left(32 \int_0^{\xb^0 \int_0^1\delb{1}\zb(\xi\xb^0,0)d\xi + z_0 } \del{\xi}\phibr(\tau,0) d\tau,0\right).
}
\end{proof}

\begin{rem} \label{vbrem}
From Lemma \ref{Jaclem}, it follows that the vacuum boundary $\Gamma_{T_0}$ in the coordinates $(\xh^0,\xh^1)$ is
given by
\eqn{vbrem1}{
\Psib(\Gamma_{T_0}) = \left\{\,
\left.\left(32 \int_0^{\xb^0 \int_0^1\delb{1}\zb(\xi\xb^0,0)d\xi + z_0 } \del{\xi}\phibr(\tau,0) d\tau,0\right) \right| \, 0\leq \xb^0 < T_0 \right\}.
}
We also note that the bounds on $\delb{1}\zb$ and $\del{\xi}\phibr(\tau,\xi)$ from Proposition \ref{eprop}.(iii) and \eqref{exist5},
respectively, imply that the map
\eqn{vbrem2a}{
\xb^0 \longmapsto 32 \int_0^{\xb^0 \int_0^1\delb{1}\zb(\xi\xb^0,0)d\xi + z_0 } \del{\xi}\phibr(\tau,0) d\tau
}
is strictly increasing. This shows that  $\Psib(\Gamma_{T_0})$ is just a reparameterization of $\Gamma_{T_0}$, and that
\eqn{vbrem2}{
\Psib(\Gamma_{T_0}) = \left\{\, (\xh^1,0) \, \left|\, 0\leq \xh^1 < \int_0^{T_0 \int_0^1\delb{1}\zb(\xi T_0,0)d\xi + z_0 } \del{\xi}\phibr(\tau,0) d\tau \, \right. \right\}.
}
This is consistent with our assertion from Section \ref{Ewave} that the vacuum boundary in the $(\xh^0,\xh^1)$ coordinates
is contained in the line $\xh^1 = 0$. This fact combined with $\wh = \sqrt{\frac{\gamma+1}{\gamma}} \delh{0}$ (see \eqref{coord1}) shows that the vacuum boundary moves with the fluid as noted previously.
\end{rem}

\subsect{exist}{Solutions with non-zero acceleration at the vacuum boundary}

With the validity of the coordinate transformation \eqref{cbar1} established,
we now turn to showing the that the maps $\{z,\zb,\psi,\phi\}$ determine a solution to the EFW equations that have non-zero acceleration
at the boundary. We begin by letting
\eqn{gwbar1}{
\gb = \Psib^* \gh = \gh_{\mu\nu}\ed \xb^\mu \ed \xb^\nu \AND \wb = \Psib^*\wh = \wb^\mu\delb{\mu}
}
denote the metric and the Frauendiener-Walton vector field.  By \eqref{coord1} and \eqref{zmet}, the
coordinate components of $\gb$ and $\wb$ are given by
\lalign{gwbar2}{
(\gb_{\mu\nu}) &= J_{\Psib}^T \begin{pmatrix}\begin{displaystyle} \frac{(1+\cos(\zb))^4}{32}\end{displaystyle} & 0\\
0 &  \begin{displaystyle}-\frac{4}{(1-\cos(\zb))^2}  \end{displaystyle} \end{pmatrix} J_{\Psib} \label{gbar2.1}
\intertext{and}
(\wb^{\mu}) & = J_{\Psi}^{-1}  \begin{pmatrix} \sqrt{2} \\ 0 \end{pmatrix} \label{gbar2.2}
}
where (see Lemma \ref{Jaclem})
\leqn{gwbar3}{
J_{\Psib} = \begin{pmatrix} \begin{displaystyle} \frac{\del{\zb}\phi(z,\zb)}{\mu(\zb)}
\end{displaystyle} & \begin{displaystyle} \frac{\del{z}\phi(z,\zb)}{\mu(\zb)} \end{displaystyle}\\
\del{z}\phi(z,\zb) & \del{\zb}\phi(z,\zb) \end{pmatrix} \begin{pmatrix}\delb{1}\zb & \delb{0}\zb \\
\delb{0}\zb & \delb{1} \zb \end{pmatrix}.
}
We also observe that the norm $\wb^2$, the square of the sound speed $\bar{s}^2$, and
the proper energy density $\rhob$ are easily computed to be
\lalign{gwbar4}{
\wb^2 & = \frac{(1-\cos(\zb))^4}{16}, \label{gwbar4.1}\\
\bar{s}^2 &= 2 \left(\frac{1-\cos(\zb)}{1+\cos(\zb)}\right), \label{gwbar4.2}
\intertext{and}
\rhob & = \frac{1}{K}\frac{1-\cos(\zb)}{1+\cos(\zb)} \label{gwbar4.3}
}
using the formulas \eqref{sound}, \eqref{cframe11}, \eqref{cframe12}, and \eqref{z2}.
From \eqref{gbar2.2} and \eqref{gwbar4.1}, we then obtain the following formula for
the fluid two-velocity
\leqn{gwvar5}{
(\vb^{\mu})  = \frac{4}{(1-\cos(\zb))^4}J_{\Psi}^{-1}  \begin{pmatrix} \sqrt{2} \\ 0 \end{pmatrix}.
}

The analysis contained in Sections \ref{eulf}, \ref{Ewave}, \ref{null}, \ref{cwave}, and
 \ref{ccord} guarantee that the pair $\{\gb_{\mu\nu},\wb^\nu\}$ defined
by the formulas \eqref{gbar2.1}, \eqref{gbar2.2}, and \eqref{gwbar3} determine a $C^2$ solution to the
EFW equations \eqref{eul1} on the spacetime region $U_{T_0,\delta}$. This, in turn, shows that
$\{\gb_{\mu\nu}, \vb^\mu, \rhob\}$, with $\rhob$ and $\vb^\mu$ given by the formulas
\eqref{gwbar4.3} and \eqref{gwvar5}, is a $C^2$ solution to the Euler equations \eqref{eul13.1}-\eqref{eul13.2}
on $U_{T_0,\delta}$.

With existence established, we are left with calculating the norm of the fluid acceleration at the
boundary. We begin by observing that
\leqn{gbarf}{
\gb = \frac{(1+\cos(\zb))^4}{32}\ed \psi \ed \psi - \frac{4}{(1-\cos(\zb))^2}
\ed\phi \ed \phi
}
and
\leqn{wbarf}{
\wb^\flat = \frac{\sqrt{2}(1+\cos(\zb))^4}{32}\ed \psi
}
where $\wb^\flat = \gb_{\mu\nu}\wb^\mu \ed \xb^\nu$.
Using \eqref{cfix2}, we see that
\leqn{dpsi1}{
\ed\psi = \frac{\del{\zb}\phi}{\mu(\zb)}\ed z + \frac{\del{z}\phi}{2\zb\mu(\zb)} \ed \zb^2,
}
and hence by \eqref{exist2}, that
\leqn{dpsi2}{
\ed\psi = \frac{32\zb^3}{\sin^3(\zb)}\del{\xi}\phibr(z,\zb^4) \ed z + \frac{4\zb^3}{\sin^3(\zb)}
\frac{\del{t}\phibr(z,\zb^4)}{\zb^4} \ed \zb^2.
}
Also, similar calculations show that
\leqn{dpsi3}{
\frac{1}{\zb^2}\ed \phi = \frac{\del{t}\phibr(z,\zb^4)}{\zb^2}\ed z + 2\del{\xi}\phibr(z,\zb^4) \ed \zb^2.
}

Next, we introduce
the dual basis
\leqn{nbasis}{
\thetab^0= \ed z \AND \thetab^1 = \ed \zb^2.
}
As we shall see, the components of the metric and Fraueniener-Walton covector field with respect
to this frame have finite limits at the vacuum boundary even though some of the $(\xb^\mu)$ coordinates components are
singular there. It is worthwhile noting that since this
basis arises from the coordinates\footnote{Recall that it was shown in Section \ref{ccord} that $(z,\zb)$
satisfies $\ed z \Wp \ed \zb = ((\delb{1}\zb)^2-(\delb{0}\zb)^2)\ed \xb^0\Wp \ed \xb^1$ with
$((\delb{1}\zb)^2-(\delb{0}\zb)^2)$ non-vanishing on $U_{T_0,\delta}$. This show that $(z,\zb)$ define
a coordinate system and it follows that $(z,\zb^2)$ does also.} $(z,\zb^2)$, we could have introduced yet one more coordinate
transformation to investigate the regularity of the fields $\{\gb,\wb\}$ near the vacuum boundary.
However, it is simpler just to work with the basis \eqref{nbasis}
without introducing another explicit coordinate transformation.

Writing the metric $\gb$ and co-vector field $\wb^\flat$ as
\eqn{gwnbasis}{
\gb = \gb_{ij}\thetab^i\thetab^j \AND \wb^\flat = \wb_i \thetab^i,
}
a straightforward calculation using \eqref{gbarf}-\eqref{nbasis} shows that
\leqn{gmatrix}{
(\gb_{ij}|_{\Gamma_{T_0}}) = \begin{pmatrix}
512 \bigl(\del{\xi}\phibr(z(\xb^0,0),0)\bigr)^2 & 64\del{\xi}\phibr(z(\xb^0,0),0)\partial^2_{t\xi}\phibr(z(\xb^0,0),0) \\
64\del{\xi}\phibr(z(\xb^0,0),0)\partial^2_{t\xi}\phibr(z(\xb^0,0),0) &
-32 \bigl(\del{\xi}\phibr(z(\xb^0,0),0)\bigr)^2 + 8\bigl( \partial^2_{t\xi}\phibr(z(\xb^0,0),0)\bigr)^2
\end{pmatrix}
}
and
\leqn{wmatrix}{
(\wb_i|_{\Gamma_{T_0}}) = \frac{1}{\sqrt{2}}\begin{pmatrix} 32\del{\xi}\phibr(z(\xb^0,0),0) &
4\partial^2_{t\xi}\phibr(z(\xb^0,0),0) \end{pmatrix}.
}
From \eqref{gmatrix}, we find that
\eqn{ginvmatrix}{
(g^{ij}|_{\Gamma_{T_0}}) = \begin{pmatrix}
\frac{4\bigl(\del{\xi}\phibr(z(\xb^0,0),0)\bigr)^2-\bigl(\partial^2_{t\xi}\phibr(z(\xb^0,0),0)\bigr)^2}
{2048\bigl(\del{\xi}\phibr(z(\xb^0,0),0)\bigr)^4} &  \frac{\partial^2_{t\xi}\phibr(z(\xb^0,0),0)}{256 \bigl(\del{\xi}\phibr(z(\xb^0,0),0)\bigr)^3}
\\
  \frac{\partial^2_{t\xi}\phibr(z(\xb^0,0),0)}{256 \bigl(\del{\xi}\phibr(z(\xb^0,0),0)\bigr)^3} &
  -\frac{1}{32 \bigl(\del{\xi}\phibr(z(\xb^0,0),0)\bigr)^2}
\end{pmatrix}.
}
Using this and \eqref{wmatrix}, we get that
\eqn{wsmatrix}{
(\wb^i|_{\Gamma_{T_0}}) = \begin{pmatrix} \begin{displaystyle} \frac{\sqrt{2}}{32\del{\xi}\phibr(z(\xb^0,0),0)}\end{displaystyle}
\\ 0 \end{pmatrix}
}
which we can use to write the frame components of $\hb$ (see \eqref{eul14}) as
\leqn{hmatrix}{
(\hb^{ij}|_{\Gamma_{T_0}}) = \begin{pmatrix}
-\frac{\bigl(\partial^2_{t\xi}\phibr(z(\xb^0,0),0)\bigr)^2}
{2048\bigl(\del{\xi}\phibr(z(\xb^0,0),0)\bigr)^4} &  \frac{\partial^2_{t\xi}\phibr(z(\xb^0,0),0)}{256 \bigl(\del{\xi}\phibr(z(\xb^0,0),0)\bigr)^3}
\\
  \frac{\partial^2_{t\xi}\phibr(z(\xb^0,0),0)}{256 \bigl(\del{\xi}\phibr(z(\xb^0,0),0)\bigr)^3} &
  -\frac{1}{32 \bigl(\del{\xi}\phibr(z(\xb^0,0),0)\bigr)^2}
\end{pmatrix}.
}
Letting
\eqn{sb2}{
\ed \bar{s}^2 = \ed \bar{s}^2_i \thetab^i,
}
we get from \eqref{gwbar4.2} that
\eqn{sb2matrix}{
(\ed\bar{s}^2_i) = \begin{pmatrix}0 & \begin{displaystyle}\frac{\sin(\zb)}{\zb}\frac{2}{(1+\cos(\zb))^2}
\end{displaystyle}\end{pmatrix}.
}
From this, we see that
\eqn{sb2matrix2}{
(\ed\bar{s}^2_i|_{\Gamma_{T_0}}) = \begin{pmatrix} 0 & \begin{displaystyle} \frac{1}{2} \end{displaystyle}\end{pmatrix}
}
which, with the help of \eqref{hmatrix}, shows that
\eqn{sbnorm}{
|\ed\bar{s}^2|_{\hb}^2|_{\Gamma_{T_0}} = \frac{1}{128 \bigl(\del{\xi}\phibr(z(\xb^0,0),0)\bigr)^2}.
}
This and \eqref{accel3} allows us to conclude that the norm of the fluid acceleration at the vacuum boundary
is non-zero and is given by the formula
\leqn{avac}{
|\bar{a}|_{\hb}|_{\Gamma_{T_0}} = \frac{1}{\sqrt{128}\del{\xi}\phibr(z(\xb^0,0),0)}.
}

We summarize the above results in the following Theorem.

\begin{thm} \label{ethm}
Suppose $k=3$, $\zb$ is the solution to the wave equation \eqref{zwave3} from Proposition \ref{eprop},
$z$ is a defined by \eqref{zdef}, $\phi$ and $\phibr$ are the maps from Proposition \ref{esprop}, and
$\psi$ is given \eqref{exist6}. Then for $T_0$ and $\delta_0$ small enough,
the triple\footnote{Here, the $\rhob$, $\vb^\mu$, and $\gb_{\mu\nu}$ are given in the $(\xb^\mu)$ coordinates with the Jacobian
matrix $J_{\Psib}$ arising from the transformation from the $(\xh^\mu)$ to the $(\xb^\mu)$ coordinates. In the $(\xh^\mu)$ coordinates,
$\rhoh$, $\vh^\mu$, and $\gh_{\mu\nu}$ are determined by \eqref{eul10}, \eqref{coord1}, and \eqref{zmet}.} $\{\gb_{\mu\nu},\rhob,\vb^\mu\}$ determined by
\alin{ethm1}{
(\gb_{\mu\nu}) &= J_{\Psib}^T \begin{pmatrix}\begin{displaystyle} \frac{(1+\cos(\zb))^4}{32}\end{displaystyle} & 0\\
0 &  \begin{displaystyle}-\frac{4}{(1-\cos(\zb))^2}  \end{displaystyle} \end{pmatrix} J_{\Psib},\\
\rhob & = \frac{1}{K}\frac{1-\cos(\zb)}{1+\cos(\zb)}\\
\intertext{and}
(\vb^{\mu}) & = \frac{4}{(1-\cos(\zb))^4}J_{\Psi}^{-1}  \begin{pmatrix} \sqrt{2} \\ 0 \end{pmatrix},
}
where
\eqn{ethm2}{
J_{\Psib} = \begin{pmatrix} \begin{displaystyle} \frac{\del{\zb}\phi(z,\zb)}{\mu(\zb)}
\end{displaystyle} & \begin{displaystyle} \frac{\del{z}\phi(z,\zb)}{\mu(\zb)} \end{displaystyle}\\
\del{z}\phi(z,\zb) & \del{\zb}\phi(z,\zb) \end{pmatrix} \begin{pmatrix}\delb{1}\zb & \delb{0}\zb \\
\delb{0}\zb & \delb{1} \zb \end{pmatrix},
}
defines a $C^2$ solution of the Euler equations \eqref{eul13.1}-\eqref{eul13.2} on the
spacetime region $U_{T_0,\delta}$. Moreover, the fluid acceleration $|\bar{a}|_{\hb}$ is
non-zero on the vacuum boundary and given by the formula
\eqn{avacA}{
|\bar{a}|_{\hb}|_{\Gamma_{T_0}} = \frac{1}{\sqrt{128}\del{\xi}\phibr(z(\xb^0,0),0)}
}
for $0\leq \xb^0 < T_0$.
\end{thm}

\sect{exact}{Exact Solutions}

In this section, we show that it is possible to determine certain classes of solutions to the Euler equations with
non-zero acceleration at the boundary that are exact in the sense they
are determined up to quadrature. First, we note that $\zb$ and $z$ given by the formulas
\eqref{eremA.1} and \eqref{zdef}, respectively, are determined up to integrals. Next, by inspection, we observe that
\leqn{phisol}{
\phi(z,\zb) = (c_1+c_2z)\left(\frac{1}{12}-\frac{1}{8}\cos(\zb) + \frac{1}{24}\cos^3(\zb)\right)
}
satisfies the wave equation \eqref{cfix3}, and in fact, satisfies the conditions \eqref{exist1.1},
\eqref{exist1.2}, and \eqref{exist1.3}.   In this case, the map $\phibr(t,\xi)$ is given by
\leqn{phibrsol}{
\phibr(t,\xi) = (c_1 + c_2 t) )\left(\frac{1}{12}-\frac{1}{8}\cos(\xi^{1/4}) + \frac{1}{24}\cos^3(\xi^{1/4})\right),
}
and it is not difficult using the powerseries expansion for $\cos(x)$ that $\phibr(t,\xi)$ admits the follow expansion
about $\xi=0$
\leqn{phibrsolexp}{
\phibr(t,\xi) = (c_1+c_2t) \left( \frac{1}{32}\xi -\frac{1}{96}\xi^{3/2}+\frac{13}{7680}\xi^2
\right) + \text{O}(\xi^{5/2}).
}
Next, using \eqref{exist6a}, it follows from \eqref{phisol} that the map $\psi$ is given by
\lalign{psisol}{
\psi(z,\zb) = & c_2\left(-\frac{1}{3}+\frac{1}{3}\ln(2) -\frac{\cos(\zb)}{3\sin^2(\zb)}+\frac{\ln\bigl(\csc(\zb)-\cot(\zb)\bigr)}{3}
 +\frac{1}{2\sin^2(\zb)}\right. \notag \\
&\text{\hspace{4.0cm}} \left. -\frac{\cot^2(\zb)}{6}-\frac{\ln(\sin(\zb))}{3}\right)
 +\left(c_1 z+\frac{c_2}{2}z^2\right). \label{psisol1.1}
}
Together, $\{z(\xb),\zb(\xb)\}$ determined by \eqref{eremA.1} and \eqref{zdef}, and  $\{\phi,\phibr,\psi\}$
determined by \eqref{phisol}, \eqref{phibrsol}, and \eqref{psisol1.1} specify completely a solution
to the Euler equations via the formulas in Theorem \ref{ethm}. Moreover, we see from \eqref{avac}
and \eqref{phibrsolexp} that the acceleration at the vaccuum boundary for these solutions is given by the formula
\eqn{aexact}{
|\bar{a}|_{\hb}|_{\Gamma_{T_0}} = \frac{32}{\sqrt{128}\bigl(c_1+c_2z(\xb^0,0)\bigr)},
}
and in particular, is constant if $c_2=0$ and time varying otherwise.

\bigskip

\noindent\emph{Acknowledgements}

\smallskip

\noindent This work was partially supported by the ARC grant DP1094582 and a MRA grant. I thank Bernd Schmidt for helpful suggestions and comments.
Part of this work was competed while visiting
the Albert-Einstein-Institute (AEI). I thank the Institute for its hospitality and for supporting this research.

\appendix

\sect{wave}{The singular wave equation $\del{t}^2\Phi + H\Phi = 0$}

Our goal in this Appendix is to prove the existence and regularity of solutions to
the singular, linear wave equation \eqref{cfix3}. To simplify notation, we will set $(z,\zb)=(t,x)$
as we will be thinking here of $z$ and $\zb$ as a time and space coordinate, respectively. Letting
\eqn{wave1}{
\Phi(t,x) = \frac{\phi(t,x)}{\sqrt{\mu(x)}},
}
we see from \eqref{cfix3} that $\Phi$ satisfies the equation
\leqn{wave2}{
\del{t}^2\Phi + H\Phi = 0
}
where $H$ is the operator
\leqn{HdefA}{
H\Phi = -\sqrt{\mu(x)}\diff{}{x}{}\left(\frac{1}{\mu(x)} \diff{}{x}{}\left(\sqrt{\mu(x)} \Phi\right) \right).
}

\subsect{fried}{The Friedrichs extension of $H$}

First, we observe that a simple integration by parts argument shows that $H$ defined \eqref{HdefA}
is symmetric on the domain $C^\infty_0(0,\pi/2)$ with respect to the standard $L^2$ inner product
\eqn{L2ip}{
\ip{\Phi_1}{\Phi_2}_{L_x^2} = \int_{0}^{\pi/2} \overline{\Phi_1(x)}{\Phi_2(x)} dx.
}
We also observe, again by integration by parts, that
\eqn{Hpos}{
H = L^\dagger L
}
where
\leqn{Ldef}{
L(\Phi) = \frac{1}{\sqrt{\mu(x)}} \diff{}{x}{}\left(\sqrt{\mu(x)}\Phi\right),
}
and $\dagger$ is the adjoint given explicitly by
\eqn{Lddef}{
L^\dagger\Phi = -\sqrt{\mu(x)} \diff{}{x}{}\left(\frac{1}{\sqrt{\mu(x)}} \Phi\right).
}

This shows that $H$ is a non-negative symmetric operator and
the quadratic form
\eqn{qdef}{
q_H(\Phi_1,\Phi_2) = \ip{\Phi_1}{H\Phi_2}
}
associated to $H$ satisfies
\eqn{qnorm1}{
q_H(\Phi_1,\Phi_2) = \ip{L\Phi_1}{L\Phi_2}
}
and in particular,
\leqn{qnorm2}{
q_H(\Phi,\Phi) = \norm{L\Phi}^2_{L^2}.
}
Defining the norm
\leqn{1norm}{
\norm{\Phi}^2_{\Hc^1}  = \norm{L\Phi}^2_{L^2} + \norm{\Phi}^2_{L^2},
}
we let
\eqn{qcomp}{
\Hc^{1} = \overline{C^\infty_0(0,\pi/2)}
}
denote the completion of $C^\infty_0(0,\pi/2)$ with respect to the norm \eqref{1norm}.
Then by Theorem X.23 of \cite{RS}, the self-adjoint Friedrichs extension of $H$, which we also denote by $H$, exists
and is defined on a dense domain
\eqn{DomH}{
D(H)\subset \Hc^{1} \subset L^2(0,\pi/2).
}

\begin{comment}
We also observe from \eqref{qdef}-\eqref{qnorm}, that
\leqn{Hker1}{
H\Phi = 0 \Longleftrightarrow  L\Phi = 0.
}
But
\eqn{Hker2}{
L\Phi = 0
}
implies that
\eqn{Hker3}{
\Phi(x) = \frac{c}{\sqrt{\mu(x)}}
}
for some $c\in \Cbb$. Since $ \frac{1}{\sqrt{\mu(x)}} \notin L^2(0,\pi/2)$, it follows from \eqref{Hker1} and \eqref{Hker3}
that
\leqn{Hker4}{
\text{Ker} H = \{0\},
}
or in other words, $(H,D(H))$ is a non-negative, injective, self-adjoint operator.
\end{comment}

\subsect{wexist}{Existence}

In order to discuss existence for the wave equation, we need define the norms
\leqn{enorm}{
\norm{\Phi}^2_{\Hc^k} = \sum_{j=0}^k \norm{H^{j/2}\Phi}^2_{L^2}   \qquad (H^{j/2} := (H^{1/2})^j)
}
and the spaces
\leqn{espace}{
\Hc^k = \bigcap_{j=0}^k D(H^{j/2}) \subset L^2(0,\pi/2),
}
where $H^{1/2}$ is the positive square root of $H$.

The following Theorem which guarantees existence and uniqueness follows from well-known results existence and
uniqueness results for abstract wave equations. See \cite{Gold,GW} for details.
\begin{thm} \label{wthm} Suppose $k\in \Nbb_0$ and  $(\Phi_0,\Phi_1)\in \Hc^{k+1}\times \Hc^k$. Then there exists a unique solution
$\Phi  \in \cap_{j=0}^{k+1} C^{j}([0,\infty),\Hc^{k+1-j})$ to the initial value problem
\alin{wthm1}{
\frac{d^2\Phi}{d t^2} + H\Phi & = 0,\\
\left(\Phi|_{t=0},\frac{d\Phi}{dt}|_{t=0}\right) &= (\Phi_0,\Phi_1).
}
\end{thm}

\subsect{reg}{Regularity near the boundary}
Away from $x=0$, the norms $\norm{\cdot}_{\Hc^k}$ are equivalent to the standard Sobolev norms and,
hence, any solution to \eqref{wave2} in $\Hc^k$ will also lie in $H^k_\text{Loc}(0,\pi/2)$.
However, the norms $\norm{\cdot}_{\Hc^k}$  are not uniformly equivalent to the standard Sobolev norms as $x$ approaches zero.
Consequently, some work is needed to
determined the regularity of functions lying in these spaces for small $x$. Our main tools to establish the
regularity will be the following:
\begin{itemize}
\item[(i)] \emph{Sobolev's inequality}: Suppose $sp < 1$, $s\in \Rbb$, and $p\in (1,\infty)$. Then
\leqn{Sob}{
\norm{\Phi}_{L^{p/(1-sp)}(a,b)}\lesssim \norm{\Phi}_{H^{s,p}(a,b)}.
}
for all $\Phi \in H^{s,p}(a,b)$.
Here, $H^{s,p}(a,b)$ denotes the fractional Sobolev spaces which coincide with standard ones for $s\in \Nbb$. We employ the usual notation $H^{s}(a,b)=H^{s,2}(a,b)$,
and we note that for  $0\leq s \leq 1$ the fractional norm $\norm{\Phi}_{H^{s}(a,b)}$ can be written as
\leqn{H1def}{
\norm{\Phi}^2_{H^s(a,b)} = \norm{\Phi}^2_{L^2(a,b)} + \int_{a}^b \int_a^b \frac{|\Phi(x)-\Phi(y)|^2}{|x-y|^{1+2s}}\,dx dy.
}
\item[(iii)]  \emph{Morrey's inequality} Suppose $p \in (1,\infty]$. Then
\leqn{Morr}{
\norm{\Phi}_{C^{0,1-1/p}} \lesssim \norm{\Phi}_{H^{1,p}(a,b)}
}
for all $\Phi \in H^{1,p}(a,b)$.
\item[(ii)] \emph{Fractional order weighted Hardy's inequality}: For $0\leq d < 1$, the following inequality
\leqn{wimbed1}{
\norm{\Phi}_{H^{1-d}(0,b)} \lesssim \norm{\Phi}_{L^2(0,b)} + \left\|x^d \diff{}{x}{\Phi }\right\|_{L^2(0,b)}
}
follows directly from the fractional order weighted Hardy's inequality \cite[Theorem 5.3]{KP} and the definition of the
fractional norm
\eqref{H1def}. We note that this type of Hardy inequality was also used in the existence results of \cite{CK1D,CK3D,JM,JM3D}.
\end{itemize}

To begin, we introduce a new coordinate
\leqn{xidef}{
\xi = \int_{0}^{x} \mu(x) dx,
}
and let
\eqn{xi0def}{
\xi_0 = \int_0^{\pi/2}\mu(s)ds.
}
We use the notation $L^2_\xi$ to denote the $L^2$ space with respect to the coordinate
$\xi$ on the interval $(0,\xi_0)$, or equivalently, the measure
\leqn{dxidef}{
d\xi = \mu(x) dx
}
on the interval $(0,\pi/2)$. We use the following notation for the $L^2$ norm:
\eqn{normdef}{
\norm{\Phi}_{L^2_\xi} = \int_{0}^{\xi_0} \Phi(\xi) d\xi.
}
We will also use the notation $H^s_\xi$ when referring to the $L^2$ Sobolev spaces
with respect to the variable $\xi$, or equivalently, with the $L^2$ spaces
defined using the differential operator
\leqn{xidiffdef}{
\diff{}{\xi}{} = \frac{1}{\mu(x)}\diff{}{x}{}
}
and the measure \eqref{dxidef}. In the following, we use the same notation to denote a function whether thought
of as a function of $x$ or a function of $\xi$ whenever this does not lead to an
ambiguity.

Since $\mu(x)$ is analytic, $\mu(x) > 0$, and $\mu(x)=\text{O}(x^3)$, it follows that
\leqn{xiprop1}{
\xi \sim x^4
}
for $x$ near zero, or equivalently
\eqn{xiprop2}{
x \sim \xi^{1/4}
}
for $\xi$ near zero. Thus, in particular,
\leqn{xiprop3}{
\mu \sim \xi^{3/4}
}
for $\xi$ near zero.

%Also, from the change of variables formula, since
%\leqn{xiprop4}{
%\left|\frac{x}{\mu(x)}\diff{}{x}{}\mu(x)\right| \leq C \quad \text{for $0<x<\pi/2$}
%}
%it follows that
%\leqn{xiprop5}{
%\left|\frac{\xi^{1/4}}{\mu(\xi)}\diff{}{\xi}{}\mu(\xi)\right| \leq C \quad \text{for $0<\xi<\xi_0$}.
%}

\begin{lem} \label{reglemB}
Suppose $\Phi \in \Hc^2$. Then
\eqn{reglemB1}{
\left\|\frac{\sqrt{\mu}\Phi}{\mu}\right\|^2_{L^2_\xi} + \left\|\diff{}{\xi}{}(\sqrt{\mu}\Phi)\right\|^2_{L^2_\xi} +
\left\|\xi^{3/4}\diff{2}{\xi}{}(\sqrt{\mu}{\Phi})\right\|^2_{L^2_\xi}
\lesssim \norm{\Phi}^2_{\Hc^2}.
}
\end{lem}
\begin{proof}
First, we observe that the identities
\leqn{Phinorm}{
|\Phi|^2 dx = \left|\frac{\Phi}{\sqrt{\mu}}\right|^2 d\xi,
}
\leqn{Lnorm}{
|L\Phi|^2 dx = \left|\diff{}{\xi}{}(\sqrt{\mu}\Phi)\right|^2 d\xi,
}
and
\leqn{Hnorm}{
|H\Phi|^2 dx = \mu^2\left|\diff{2}{\xi}{}(\sqrt{\mu}\Phi)\right|^2 d\xi
}
follow directly from the definitions \eqref{HdefA}, \eqref{Ldef}, \eqref{xidef}, \eqref{dxidef}, and
\eqref{xidiffdef}. Integrating \eqref{Phinorm}, \eqref{Lnorm}, and \eqref{Hnorm} then gives
\leqn{reglemB2}{
\left\|\frac{\sqrt{\mu}\Phi}{\mu}\right\|^2_{L^2_\xi} + \left\|\diff{}{\xi}{}(\sqrt{\mu}\Phi)\right\|^2_{L^2_\xi} +
\left\|\xi^{3/4}\diff{2}{\xi}{}(\sqrt{\mu}{\Phi})\right\|^2_{L^2_\xi}
\lesssim \left\|\Phi\right\|^2 + \norm{L\Phi}^2_{L^2}
+ \left\|H\Phi\right\|^2.
}
Since
\eqn{reglemB3}{
 \norm{L\Phi}^2_{L^2}  = \norm{H^{1/2}\Phi}^2_{L^2}
}
by \eqref{qnorm2}, the proof follows from \eqref{reglemB2}.
\end{proof}

\begin{lem} \label{reglemC}
Suppose $\Phi \in \Hc^4$. Then
\eqn{reglemC1}{
\norm{\sqrt{\mu}\Phi}_{C^{1,1/2}_\xi} \lesssim \norm{\Phi}_{\Hc^4},
}
and
\eqn{reglemC2}{
|\sqrt{\mu(\xi)}\Phi(\xi)|\lesssim \norm{\Phi}_{\Hc^4}\xi.
}
\
\end{lem}
\begin{proof}
First, we observe that
\lalign{reglemC2a}{
\left\|\frac{\mu^2}{\xi^{3/4}}\diff{2}{\xi}{}(\sqrt{\mu}\Phi)\right\|_{L^2_\xi}
&\lesssim \left\|\xi^{3/4}\diff{2}{\xi}{}(\sqrt{\mu}\Phi)\right\|_{L^2_\xi} && \text{(by \eqref{xiprop3})}\notag\\
&\lesssim \norm{\Phi}_{\Hc^2} && \text{(by Lemma \ref{reglemB})}. \label{reglemC2a.1}
}
Next, we see from \eqref{HdefA} that
\eqn{reglemC3}{
|H^2\Phi|^2 dx = \mu^2\left| \diff{2}{\xi}{} \left(\mu^2\diff{2}{\xi}{}(\sqrt{\mu}\Phi)\right)\right|^2 d\xi,
}
and so upon integrating, we find, after using \eqref{xiprop3}, that
\leqn{reglemC4}{
\left\|\xi^{3/4} \diff{2}{\xi}{} \left(\mu^2\diff{2}{\xi}{}(\sqrt{\mu}\Phi)\right)\right\|^2_{L^2_\xi} \lesssim
\norm{\Phi}^2_{\Hc^4}.
}
Setting
\eqn{reglemC5}{
f = \mu^2\diff{2}{\xi}{}(\sqrt{\mu}\Phi),
}
the two inequalities \eqref{reglemC2a.1} and \eqref{reglemC4} show that
\leqn{reflemC6}{
\left\|\frac{1}{\xi^{3/4}} f\right\|^2_{L^2_\xi} + \left\|\xi^{3/4}\diff{2}{\xi}{f}\right\|^2_{L^2_\xi} \lesssim \norm{\Phi}^2_{\Hc^4}.
}
Letting
\eqn{reflemC7}{
f_\lambda(\xi) = f(\lambda\xi),
}
we compute
\alin{reflemC8}{
\Bigl|\sup_{\lambda/2 < \xi  < \lambda} |f(\xi)|\Bigr|^2 & =
\Bigl|\sup_{1/2 < \xi  < 1} |f_\lambda(\xi)|\Bigr|^2 \\
& \lesssim \norm{f_\lambda}^2_{H^1(1,2)}  && \text{(by \eqref{Morr})}\\
&  \lesssim \int_{1/2}^1 |f_\lambda(\xi)|^2d\xi + \int_{1/2}^1 \left|\diff{2}{\xi}{f_\lambda}\right|^2 d\xi \\
&  \lesssim \int_{1/2}^1 \frac{1}{\xi^{3/2}}|f(\lambda\xi)|^2 d\xi +
\int_{1/2}^1 \xi^{3/2}\lambda^4 \left|\diff{2}{\xi}{f}(\lambda\xi)\right|^2 d\xi  \\
& = \lambda^{1/2}\int_{\lambda/2}^{\lambda} \frac{1}{\xi^{3/2}}|f(\xi)|^2 d\xi +
\lambda^{3/2}\int_{\lambda/2}^{\lambda} \xi^{3/2}\left|\diff{2}{\xi}{f}(\xi)\right|^2 d\xi  \\
& \lesssim \lambda^{1/2}\norm{\Phi}^2_{\Hc^4},
}
where in deriving the lass inequality we used \eqref{reflemC6}).
From this, we conclude that
\leqn{reflemC9}{
|f(\xi)| \lesssim \norm{\Phi}_{\Hc^4}|\xi|^{1/4}.
}
In particular, this shows that
\leqn{reflemC10}{
\lim_{\xi \searrow 0} f(\xi) = 0.
}

Next, we see that from \eqref{reflemC6} and the inequality \eqref{wimbed1} that
\eqn{reflemC11}{
\norm{f}_{H^{5/4}_\xi} \lesssim \norm{\Phi}_{\Hc^4}.
}
But since
\eqn{reflemC12}{
\norm{f}_{C^{0,3/4}_\xi} \lesssim \norm{f}_{H^{1,4}_\xi}\lesssim \norm{f}_{H^{5/4}_\xi}
}
by \eqref{Sob} and \eqref{Morr}, we see that
\eqn{reflemC13}{
\norm{f}_{C^{0,3/4}_\xi}  \lesssim \norm{\Phi}_{\Hc^4}.
}
This together with \eqref{reflemC10} shows that
\eqn{reflemC14a}{
|f(\xi)| \lesssim \norm{\Phi}_{\Hc^4}|\xi|^{3/4},
}
which in turn implies that
\eqn{reflemcC15}{
\left|\xi^{1/4+\epsilon}\diff{2}{\xi}{}(\sqrt{\mu}\Phi)\right|\lesssim \left|\frac{\xi^{1/4+\epsilon}}{\mu^2(\xi)}f(\xi)\right|
\lesssim \frac{\norm{\Phi}_{\Hc^4}}{|\xi|^{1/2-\ep}}
}
for any $\ep>0$. Integrating, we find that
\leqn{reflemC16}{
\left\|\xi^{1/4+\epsilon}\diff{2}{\xi}{}(\sqrt{\mu}\Phi)\right\|^2_{L^2_\xi}\lesssim
\norm{\Phi}_{\Hc^4}^2,
}
and this inequality combined with Lemma \ref{reglemB} and the inequality \eqref{wimbed1}
shows that
\eqn{reflemC17}{
\left\|\diff{}{\xi}{}(\sqrt{\mu}\Phi)\right\|_{H^{3/4+\epsilon}_\xi} \lesssim \norm{\Phi}_{\Hc^4}^2.
}
Applying Morrey's inequality \eqref{Morr}, we arrive at
\eqn{reflemC18}{
\left\|\diff{}{\xi}{}(\sqrt{\mu}\Phi)\right\|_{C^{0,1/4}_\xi} \lesssim \norm{\Phi}_{\Hc^4}.
}

Next, we observe that
\eqn{reflemC19}{
\left\|\frac{1}{\xi^{3/4}}\sqrt{\mu}\Phi\right\|_{L^2_\xi}
+ \left\|\xi^{3/4}\diff{2}{\xi}{}\bigl(\sqrt{\mu} \Phi\bigr) \right\|_{L^2_\xi} \lesssim \norm{\Phi}^2_{\Hc^2}
}
by Lemma \ref{reglemB} and \eqref{xiprop3}. The same argument used to derive \eqref{reflemC9}
from \eqref{reflemC6} shows that
\leqn{reflemC20}{
|\sqrt{\mu(\xi)}\Phi(\xi)|\lesssim \norm{\Phi}_{\Hc^2}\xi^{1/4}.
}

Together, Lemma \ref{reglemB}, the inequality \eqref{reflemC16}, and \eqref{wimbed1} show that
\eqn{reflemC21}{
\norm{\sqrt{\mu}\Phi}_{H^{7/4-\epsilon}_\xi} \lesssim \norm{\Phi}_{\Hc^4},
}
and hence that
\leqn{reflemC22}{
\norm{\sqrt{\mu}\Phi}_{C^{1,1/2}_\xi} \lesssim \norm{\Phi}_{\Hc^4}
}
by Morrey's inequality \eqref{Morr}.  Finally, since
\eqn{reflemC23}{
\lim_{\xi \searrow 0} \sqrt{\mu(\xi)}\Phi(\xi) = 0
}
by \eqref{reflemC20}, we get from \eqref{reflemC22} that
\eqn{reflemC14}{
|\sqrt{\mu(\xi)}\Phi(\xi)| \lesssim \norm{\Phi}_{\Hc^4}\xi.
}
\end{proof}

\begin{thm} \label{regthm}
Suppose $(\Phi_0,\Phi_1) \in \Hc^6\times \Hc^5$ and
$\Phi  \in \cap_{j=0}^{6} C^{j}([0,\infty),\Hc^{6-j})$
is the solution to
the wave equation \eqref{wave2} from Theorem \ref{wthm}. Then
there exists a map $\phibr(t,\xi)$ that satisfies the following:
\begin{itemize}
\item[(i)]
\eqn{regthm1}{\phibr \in C^1\bigl([0,\infty),C^{1,1/2}(0,(\pi/2)^4)\bigr)\cap \bigcap^5_{j=0} C^{j}\bigl([0,\infty),C^{5-j}(0,(\pi/2)^4)\bigr),
}
\item[(ii)] for any $t\in (0,\infty)$,
\eqn{regthm2}{|\del{t}\phibr(t,\xi)| + |\phibr(t,\xi)| \lesssim \xi}
for  $0<\xi < (\pi/2)^4$, and
\item[(iii)]
\eqn{regthm3}{
\sqrt{\mu(x)}\Phi(t,x) = \phibr(t,x^4)
}
for all $(t,x)\in (0,\infty)\times (0,\pi/2)$.
\end{itemize}
Moreover, if there exists a positive constant $c$ such that
\eqn{regthm4}{
\frac{1}{\mu(x)}\del{x}\bigl(\sqrt{\mu(x)}\Phi(0,x)\bigr) \geq c > 0
}
for all $x\in (0,\pi/2)$, then there exists a $T>0$ such that
\eqn{regthm5}{
\del{\xi}\phibr(t,\xi) > 0  %\frac{4x^3}{\mu(x)} \del{\xi}\Phibr(t,x^4) > 0
}
for all $(t,x) \in (0,T)\times (0,(\pi/2)^4)$.
\end{thm}
\begin{proof}
Statements (i)-(iii) follows directly from the regularity statement $\Phi  \in \cap_{j=0}^{6} C^{j}([0,\infty),\Hc^{6-j})$,
Lemma \eqref{reglemC},  and that fact that
$C^{\ell}\subset \Hc^{\ell+1}$ which follows from Morrey's inequality and the inclusion $\Hc^\ell\subset H^\ell_{\text{Loc}}(0,\pi)$.
Note that we are also using the fact that the variable $\xi$ is uniformly equivalent to $x^4$, see \eqref{xiprop1}.

For the final statement, suppose that
\leqn{regthm6}{
\frac{1}{\mu(x)}\del{x}\bigl(\sqrt{\mu(x)}\Phi(0,x)\bigr) \geq c > 0
}
for $x\in (0,\pi/2)$.  Then differentiating $\sqrt{\mu(x)}\Phi(t,x) = \phibr(t,x^4)$ gives
\leqn{regthm7}{
\frac{1}{\mu(x)} \del{x} \bigl(\sqrt{\mu(x)}\Phi(t,x)\bigr) = \frac{4 x^3}{\mu(x)}\del{\xi}\phibr(t,x^4).
}
Since there exists a non-zero constant $C$ such that
\eqn{regthm8}{
0 < \frac{1}{C} \leq \frac{4 x^3}{\mu(x)} \leq C
}
for $0<x<\pi/2$, it follows from \eqref{regthm6} and \eqref{regthm7} that
\eqn{regthm9a}{
\del{\xi}\phibr(0,x^4) \geq \frac{c}{C} > 0
}
for all $x\in (0,\pi/2)$, or equivalently
\eqn{regthm9b}{
\del{\xi}\phibr(0,\xi) \geq \frac{c}{C} > 0
}
for all $\xi \in (0,(\pi/2)^4)$. From the continuity of $\del{\xi}\phi(t,\xi)$, we see that there exists
a $T>0$ such that
\eqn{regthm9}{
\del{\xi}\phibr(t,\xi)  \geq \frac{c}{2C} > 0
}
for all $(t,\xi) \in [0,T)\times (0,(\pi/2)^4)$.
\end{proof}

%\end{spacing}

\end{document}